\documentstyle[proceedings,psfig]{crckapb} 

\begin{opening}
\title{Models and Observations of the Chemistry near Young Stellar
Objects}

\author{Ewine F.\ van Dishoeck}

\institute{Leiden Observatory
           P.O.\ Box 9513\\
           2300 RA Leiden,
           The Netherlands \\
           e-mail: ewine@strw.leidenuniv.nl}

\author{Michiel R.\ Hogerheijde}

\institute{Astronomy Department,  
           Univ.\ of California \\
           Berkeley, CA 94720, USA\\
           e-mail: michiel@astro.berkeley.edu}

\end{opening}

\runningtitle{Chemistry around YSOs}

\begin{document}

\section{Introduction}

\begin{figure}[tb]
\begin{center}
\leavevmode
\psfig{figure=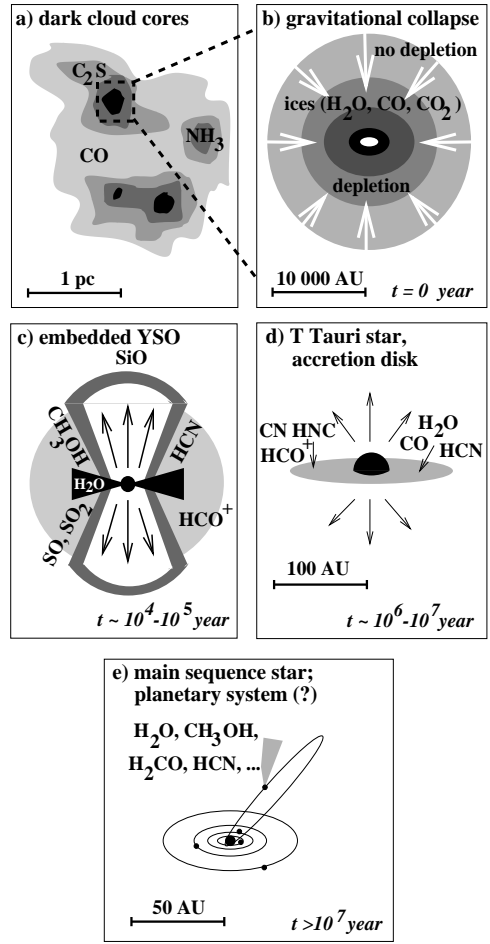,height=12.5cm,angle=0}
\end{center}
\caption{A schematic view of the characteristic molecules at different
stages of low-mass star formation.  (a) Dark cloud cores, where
radicals and unsaturated carbon chains are prominent; (b) Collapse
stage, where high levels of depletion of molecules have been inferred;
(c) Deeply embedded YSO phase, where heating and supersonic outflows
result in evaporation of ices and create a variety of chemical
environments; (d) Young T Tauri star with a residual protoplanetary
accretion disk containing gas-phase molecules and ices; (e) Mature
planetary system with icy bodies such as comets (Figure by M.R.\
Hogerheijde, after Shu et al.\ 1987; from: van Dishoeck \& Blake 1998).
}
\label{fig-1}
\end{figure}

The study of the chemical evolution of gas and dust from pre-stellar
dense cores to circumstellar disks around young stars forms an
essential part of understanding star- and planet formation.
Throughout the collapse- and protostellar phases, simple and complex
molecules are formed, many of which deplete onto cold grains and are
eventually incorporated into the icy planetesimals of new solar
systems (see Figure~\ref{fig-1}).  Tracing this chemical evolution
provides a wealth of information, not only about the chemical
processing in primitive solar nebulae, but also about physical
processes which occur in the immediate surroundings of young stellar
objects (YSOs).

\begin{figure}[tb]
\vspace{-1.7cm}
\begin{center}
\leavevmode
\psfig{figure=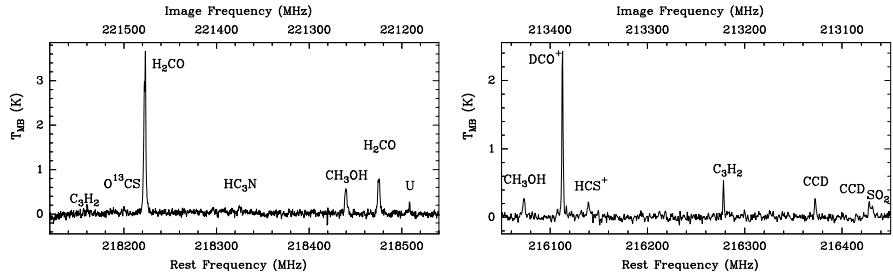,width=14cm,angle=90}
\end{center}
\vspace{-4.5cm}
\caption{JCMT submilllimeter spectra of the deeply embedded low-mass
YSO IRAS 16293$-$2422, showing lines of characteristic molecules
such as H$_2$CO, CH$_3$OH  and deuterated species
(from: van Dishoeck et al.\ 1995).}
\label{fig-2}
\end{figure}

Interstellar chemistry has been an active field of research for more
than 60 years, ever since the detection of the first molecules in
1937--1941. Nearly 120 different gas-phase species have been
identified (see Table~\ref{tab-mol}), not including isotopic
varieties, with abundances down to $10^{-11}$ with respect to
H$_2$. The majority of the species have been detected through their
rotational transitions at millimeter wavelengths.  Most of the early
observations were performed with typical beam sizes of $1'$,
corresponding to linear scales of nearly 10,000 AU (0.04 pc) in the
nearest star-forming regions in Taurus and Ophiuchus. With the advent
of large single-dish submillimeter telescopes and millimeter
interferometers, it has become possible to study the dense envelopes
around YSOs more directly on scales of $\sim 2''$--$20''$ (300--3000 AU
at 150 pc) (Blake 1997) (see Figure~\ref{fig-2}).

Although much effort has  focused on gas-phase chemistry, water
ice was identified already in 1973 through its vibrational absorption
toward bright infrared sources. The enormous advances in infrared
instrumentation in the last decade, both from the ground and from
space with the {\it Infrared Space Observatory} (ISO), have resulted
in the detection of several other solid-state species and allow a
complete inventory to be made (Tielens \& Whittet 1997).  The
development of realistic models which include both gas-phase and
grain-surface chemistry provides a challenge to theorists.

The combination of new observations and models has lead to the
following scenario for the chemistry during star formation (see
Figure~\ref{fig-1}).  In the cold molecular cores prior to star
formation, the chemistry is dominated by low-temperature gas-phase
reactions leading to the enhanced formation of small radicals and
unsaturated molecules. Long carbon-chains are produced if the gas is
initially atomic-carbon rich.  During the cold pre-stellar and
collapse phases, many molecules accrete onto the grains and form an
icy mantle. Here surface chemistry and processing by ultraviolet
photons, X-rays and cosmic rays can modify the composition.  Once the
new star starts to warm the surrounding envelope, the ices are heated
and molecules evaporate back into the gas phase, probably in a
sequence according to their sublimation temperatures. In addition, the
outflows from the young star penetrate the envelope, creating high
temperature shocks and lower temperature turbulent regions in which
both volatile and refractory material containing silicon can be
returned. These freshly evaporated molecules then drive a rich
chemistry in the `hot cores' for a period of $10^5$ yr. Finally, the
envelope is dispersed by winds and/or ultraviolet photons, leading to
the appearance of photon-dominated regions.

Part of the gas-phase and icy material is incorporated into the
circumstellar disks, where they survive for a few $\times 10^{6}$ yr
before being dispersed or assembled into new planetary bodies.
Indeed, the connection between interstellar chemistry and solar system
material has been greatly strengthened by observations of comets
Hyakutake and Hale--Bopp, which reveal a remarkable similarity between the
composition of interstellar and cometary ices (e.g., Ehrenfreund et
al.\ 1997, Irvine et al.\ 1999).

\begin{table}[tb]
\begin{center}
\caption{Identified interstellar and circumstellar gas-phase molecules}
\label{tab-mol}
\vspace{-2.2cm}
\leavevmode
\psfig{figure=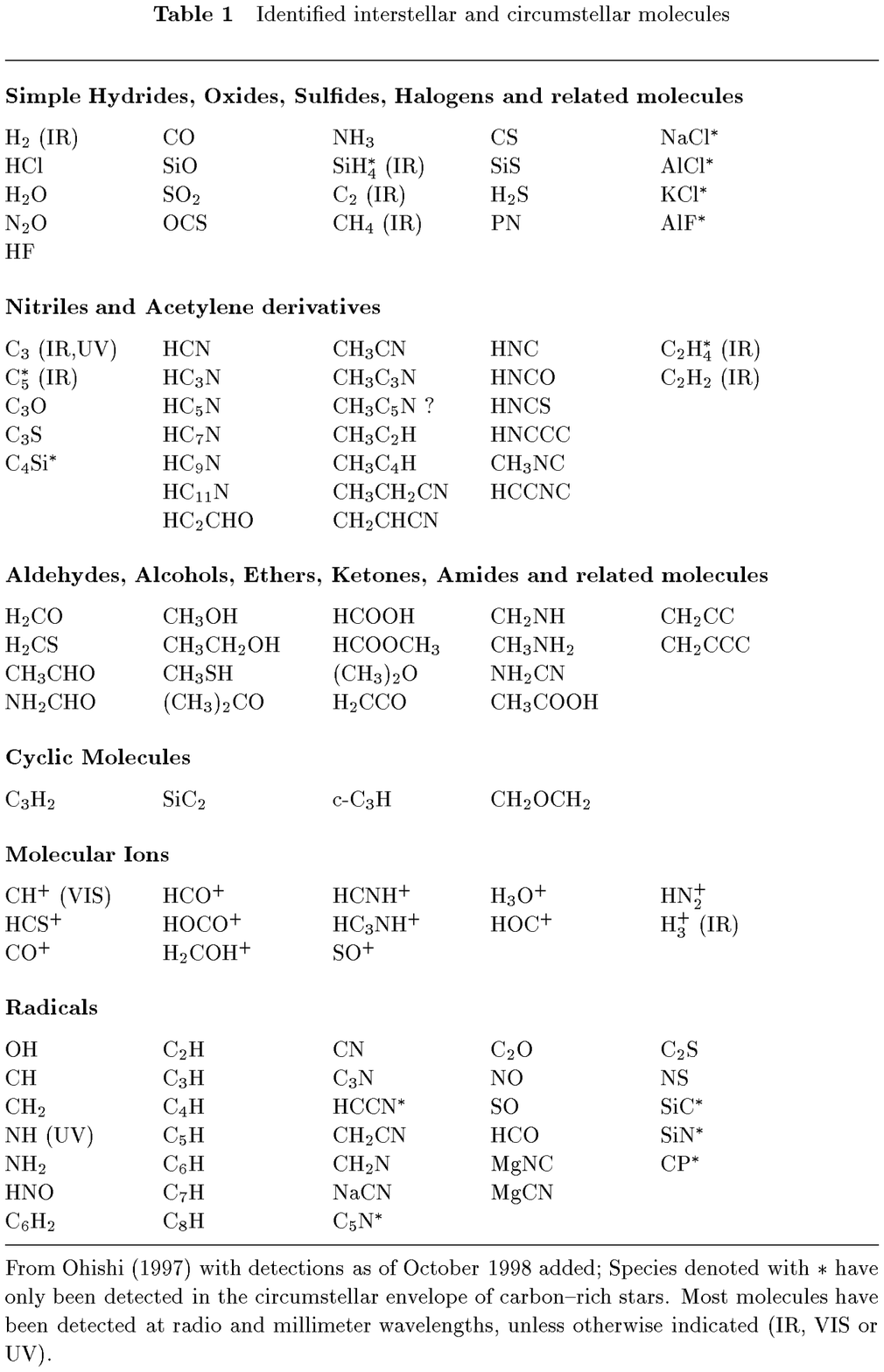,width=15.0cm}
\end{center}
\vspace{-4.5 true cm}
\end{table}

Molecular spectroscopy is also a very powerful tool to probe the
enormous range in temperatures from 10 to 10,000~K and densities from
$10^4$ to $10^{13}$ cm$^{-3}$ that accompany the star-formation
process. In addition, molecular lines provide the only means to trace
the velocity fields in the highly-extincted circumstellar environment.
In fact, the two analyses go hand in hand.  On the one hand, an
accurate physical model is a prerequisite for deriving reliable
molecular abundances.  On the other hand, it is important to know
which molecule traces which physical component if its emission is to be used
as a probe of the physical parameters.

This chapter first discusses the basic physical and chemical processes
that play a role during star formation.  Subsequently, the
observational methods and problems in deriving of abundances are
discussed. Finally, recent results on observations and models of the
chemistry in the various star-formation phases are presented.  This
discussion focuses on those species whose abundances are particularly
enhanced or decreased at a certain phase. These species often have
only minor abundances in terms of overall composition ---less than
$10^{-7}$ with respect of hydrogen---, but provide important
`signposts' or `clocks' of the evolutionary state of an object.
This chapter is based on the recent reviews by van Dishoeck (1998a) on
the chemistry in diffuse and dense clouds, and by van Dishoeck \&
Blake (1998), van Dishoeck (1998b) and Langer et al.\ (1999) on the
chemistry in star-forming regions. Other reviews on this subject
include Hartquist et al.\ (1998), Irvine (1998) and van Dishoeck et
al.\ (1993).  For an excellent overview of the physics of molecular
clouds, see Genzel (1992).

\section{Gas-phase chemistry}

\subsection{Basic molecular processes}

The networks developed to describe the gas-phase chemistry in
star-forming regions contain thousands of reactions between several
hundreds of species (Millar et al.\ 1997a, Lee et al.\ 1996a). These
thousands of reactions involve only a handful of different basic
molecular processes, however, which have been described in detail in
reviews by Dalgarno (1987) and van Dishoeck (1988).  At the low
densities in the interstellar gas, three-body processes are
unimportant, so that only two-body reactions need to be considered.
The rate of a reaction between species $X$ and $Y$ is given by $k
n$(X)$n$(Y) in cm$^{-3}$ s$^{-1}$, where $k$ is the reaction rate
coefficient in cm$^3$ s$^{-1}$ and $n$ is the concentration in
cm$^{-3}$.

There are two basic mechanisms by which molecular bonds can be formed
(see Table~\ref{tab-proc}).  The first process is radiative
association of gas-phase atoms or molecules, in which the new molecule
is stabilized by the emission of a photon. The second mechanism is the
formation of molecules on the surfaces of grains, in which the grain
carries off the released energy corresponding to the molecular bond.
The rate coefficients for the formation processes range from less than
$10^{-17}$ up to $10^{-13}$ cm$^3$ s$^{-1}$ and have considerable
uncertainties, often up to an order of magnitude.

\begin{table}
\begin{center}
\caption{Some basic microscopic processes}
\label{tab-proc}
\begin{tabular}{ll}
\hline
Process & Name \\
\hline
\\ [0pt]
{\it Chemical Processes} \\ [5pt]
X + Y $\to$ XY + h$\nu$ & Radiative association \\
X + Y:grain $\to$ XY + grain & Grain-surface formation \\
XY + h$\nu$ $\to$ X + Y & Photodissociation \\
XY$^+$ + $e$ $\to$ X + Y & Dissociative recombination \\
X$^+$ + YZ $\to$ XY$^+$ + Z & Ion--molecule reaction\\
X$^+$ + YZ $\to$ X + YZ$^+$ & Charge--transfer reaction\\
X + YZ $\to$ XY + Z     & Neutral--neutral reaction \\ [10pt]

{\it Heating and Cooling Processes} \\ [5pt]

grain or PAH + h$\nu$ $\to$ grain$^+$ or PAH$^+$ + $e^*$
      & Photoelectric heating \\
H$_2$ + cosmic ray $\to$ H$_2^+$ + $e^*$ & Cosmic ray heating \\
CO($J$) + coll. $\to$ CO($J^*$) $\to$ CO($J'$) + h$\nu$ & CO line cooling \\
O($^3$P$_2$) + coll. $\to$ O($^3$P$_1$) $\to$ O($^3$P$_2$) + h$\nu$
   & [O I] line cooling \\
C$^+$($^2$P$_{1/2}$) + coll. $\to$ C$^+$($^2$P$_{3/2}$) $\to$
C$^+$($^2$P$_{1/2}$) + h$\nu$ & [C II] line cooling \\
gas + grain $\to$ gas$'$ + grain$'$ & Gas--grain heating or cooling \\ 

\hline
\end{tabular}
\end{center}
\end{table}

Molecules are readily destroyed by the absorption of ultraviolet
photons.  This process is very effective at the edges of dark clouds
and in clouds located close to young stars (so-called photon-dominated
regions (PDRs), Hollenbach \& Tielens 1997, 1999). For molecules such
as H$_2$ and CO, the photodissociation can take place only at very
short wavelengths between 912 and 1100 \AA, whereas most other species
are dissociated by radiation out to 3000~\AA.  The ultraviolet photons
can also ionize atoms with ionization potentials less than 13.6~eV
(e.g., C, S, Si, Fe, Mg, ...), thereby increasing the electron
abundance inside the cloud.  In the unshielded interstellar radiation
field such as given by Draine (1978), typical rates are
$10^{-9}$--$10^{-10}$ s$^{-1}$, corresponding to lifetimes against
photodissociation or photoionization of only $10^2$--$10^3$ yr.

Inside a cloud, the ultraviolet radiation is reduced because of
absorption and scattering by grains (e.g., Roberge et al.\ 1991).
Deep inside dark clouds ($> 5$ mag), little of the ambient radiation
penetrates, but a weak ultraviolet field is maintained by cosmic-ray
induced photons, resulting from the excitation of H$_2$ by secondary
electrons produced by cosmic ray ionization of H$_2$ (Gredel et al.\
1989). For a typical cosmic ray ionization rate $\zeta$ of a few
$\times 10^{-17}$ s$^{-1}$, the resulting photorates are four to five
orders of magnitude lower than those in the unshielded interstellar
radiation field at the edge.

Molecular ions are efficiently destroyed by the process of
dissociative recombination. Rate coefficients in 10--30 K gas are
typically $10^{-7}$ to $10^{-6}$ cm$^3$ s$^{-1}$, with the lower
values representative for H$_3^+$ ---a key species in the chemistry
(Dalgarno 1994, Larsson 1997).  A major uncertainty in the models is
the branching ratio to the various products. Complete experimental
data are only just becoming available, and recent results indicate
that three-body product channels (e.g., H$_3$O$^+$ + e $\to$ OH + H +
H or O + H$_2$ + H) have a much larger probability than thought
previously (e.g., Vejby-Christensen et al.\ 1997).

\begin{figure}[tb]
\vspace{-1.5cm}
\begin{center}
\leavevmode
\psfig{figure=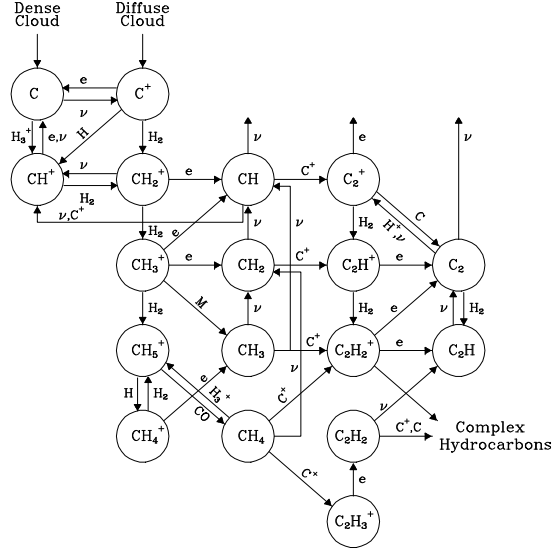,width=17cm,angle=-90}
\end{center}
\vspace{-3.2cm}
\caption{Initial steps in the gas-phase carbon chemistry in diffuse and
dark clouds.}
\label{fig-3}
\end{figure}

In dense clouds, destruction of neutral molecules can also occur
through chemical reactions, in particular with He$^+$ (e.g., He$^+$ +
CO $\to$ He + C$^+$ + O).  Collisional dissociation is only important
in regions of very high temperature ($>3000$ K) and density, such as
found in shocks.

Once molecular bonds have been formed, they can be rearranged by
chemical reactions leading to more complex species. The first
generation of models (e.g., Herbst \& Klemperer 1973, Prasad \&
Huntress 1980) included primarily ion-molecule reactions. The majority
of these reactions are known to be fast at low temperatures owing to
the long-range attraction between the ion and the molecule.  It has
recently been realized that neutral--neutral reactions involving
radicals and/or atoms can also have rate coefficients as large as
$\sim 10^{-10}$--$10^{-9}$ cm$^{3}$ s$^{-1}$ at low temperatures (e.g.,
Smith 1997, Kaiser et al.\ 1997).

\subsection{Gas-phase networks}

The gas-phase chemical networks describing the most important
formation routes of various molecules based on the above processes
have been discussed extensively in the literature, see reviews by,
e.g., Millar (1990), Herbst (1995) and van Dishoeck (1998a).  The
abundances of the elements play a key role in the chemistry,
especially the absolute and relative amounts of gas-phase carbon and
oxygen in atomic or molecular form.  It has recently become clear that
the interstellar abundances of most elements are $\sim30$\% lower than
the solar abundances (see Meyer 1997 for an overview). The fraction of
heavy elements in the gas phase is large (at least 30\%) in diffuse
clouds, except for the most refractory elements (Si, Fe, ..), which
are 99\% in solid form.  For dense clouds, the depletions of species
like C, N, O and S are expected to be more significant, but are ill
constrained observationally.

Because hydrogen is so much more abundant than any other element,
reactions with H and H$_2$ dominate the networks if they are
exothermic.  This is only the case for small ions at low temperatures;
most reactions of neutrals and large ions with H or H$_2$ have
substantial barriers. H$_2$ itself is produced predominantly on the
surfaces of grains (Hollenbach \& Salpeter 1971), but once formed it
is essential for the gas-phase chemistry of other species.  Ionization
of H$_2$ by cosmic rays at a rate $\zeta$ leads to H$_2^+$, which
reacts with H$_2$ to form the stable H$_3^+$ ion. This ion plays a
pivotal role in the subsequent ion-molecule chemistry through proton
transfer.  Its recent detection in interstellar space provides strong
observational support for the gas-phase chemical networks (Geballe \&
Oka 1996, McCall et al.\ 1998).

The production of complex hydrocarbons in dense clouds occurs via
three types of pathways (Herbst 1995): (i) carbon insertion reactions
(e.g. C$^+$ + CH$_4$ $\to$ C$_2$H$_2^+$ + H$_2$ or C + C$_2$H$_2$ $\to$ C$_3$H
+ H); (ii) condensation reactions (e.g. CH$_3^+$ + CH$_4$ $\to$
C$_2$H$_5^+$ + H$_2$ or C$_2$H + C$_2$H$_2$ $\to$ C$_4$H$_2$ + H); and
(iii) radiative association reactions (e.g. C$^+$ + C$_n$ $\to$
C$_{n+1}^+$ + h$\nu$) (see also Figure~\ref{fig-3}).  In general,
carbon insertion with C$^+$ is thought to be the dominant route. Since
this leads to loss of one hydrogen and because the larger ions
C$_n$H$_m^+$ do not react rapidly with H$_2$, low-temperature
gas-phase chemistry produces strongly unsaturated hydrocarbons, in
agreement with observations of dark clouds. The necessary C$^+$ ions
are produced in small amounts through destruction of CO by He$^+$.
Reactions with C may be competitive if they are as rapid as suggested
by recent laboratory experiments.  Once carbon is locked up in the
very stable CO molecule, the formation of more complex hydrocarbons
ceases.

At high temperatures of $\sim 200$--$2000$~K such as encountered in the
`hot core' regions surrounding massive young stars and in shocks,
gas-phase reactions with H and H$_2$ become significant.
Specifically, the reactions 
$ \rm O \ + \ H_2 \ \rightarrow \ \rm OH \ + H \ - 4480\ K $
and 
$ \rm OH \ + \ H_2 \ \rightarrow \ \rm H_2O \ + H - 2100 \ K $ 
start to proceed at temperatures of a few hundred K (Ceccarelli et
al.\ 1996, Charnley 1997), and drive most of the available gas-phase
oxygen not locked up in CO into H$_2$O. Similar reactions of S with
H$_2$ convert the gas-phase sulfur into H$_2$S.  However, the back
reactions of OH and H$_2$O with H have comparable energy barriers, so
that the balance between O, OH and H$_2$O depends also on the H/H$_2$
ratio in the warm gas.

When most of the oxygen has been driven into H$_2$O at high
temperatures, little O, OH and O$_2$ are available in the gas,
preventing the formation of, for example, SO$_2$. Since O and O$_2$
also destroy reactive species such as CN, the CN abundance is enhanced
at higher temperatures leading to enhanced formation of HC$_3$N
through reaction with C$_2$H$_2$.

\subsection{Gas-phase models}

\subsubsection{Depth-dependent vs.\ time-dependent models}

The calculation of the abundances in star-forming regions requires a
physical model in which the temperature, density, radiation field
etc.\ are specified as functions of position and/or time.  In general,
two different classes of models are considered: (i) Steady-state,
depth-dependent models, in which the physical parameters and
molecular abundances do not change with time, but are functions of
depth into the cloud. Models of the translucent outer envelopes of
clouds (e.g., van Dishoeck 1998a) and of dense ultraviolet photon- or
X-ray dominated regions (PDRs or XDRs, see e.g., Hollenbach \& Tielens
1997, Sternberg et al.\ 1997, Maloney et al.\ 1996) fall in this
category.  (ii) Time-dependent, depth-independent models, in which
the concentrations are computed as functions of time at a single
position deep into the cloud. Models of dark pre-stellar cores (e.g.,
Lee et al.\ 1996a, Millar et al.\ 1997b), collapsing envelopes (e.g.,
Bergin \& Langer 1997) and hot cores near massive young stars (e.g.,
Charnley et al.\ 1992) fall into this category. The time scale for
reaching chemical equilibrium depends on the density, temperature and
ionization fraction and is typically $10^5$--$10^7$~yr.

In both cases, the parameters that enter the models are (1) the
elemental abundances of C, O, N, S, metals...; and (2) the cosmic ray
ionization rate $\zeta$ in s$^{-1}$.  In the steady-state,
depth-dependent models, additional parameters are (3) the geometry
(e.g. plane-parallel, spherical, ...); (4) the density $n_{\rm
H}$=$n$(H) + 2$n$(H$_2$) as a function of position; (5) the incident
radiation field, specified by a factor $I_{UV}$ times the standard
interstellar radiation field as given by, e.g., Draine (1978); and (6)
the grain parameters, i.e., the extinction curve, albedo and
scattering function.  The temperatures of the gas and dust as
functions of position in the cloud can be obtained self-consistently
from the balance of the heating and cooling processes listed in
Table~\ref{tab-proc}. Alternatively, they can be constrained from
observations and provided as additional input parameters (see \S 4.2).

In the time-dependent models, the additional parameters besides (1)
and (2) are: (3) the density, usually taken to be constant with time
(so-called pseudo time-dependent models); (4) the visual extinction
$A_V$ at the position in the cloud, usually taken to be so large that
external photons can be neglected; and (5) the initial abundances of
the various species at $t$=0, often taken to be atomic except for
H$_2$. The temperature can be obtained from the thermal balance, but
is almost always set at 10~K for both the gas and dust, typical of a
dark cloud shielded from ultraviolet radiation and heated by cosmic
rays only.  In comparison with observations, the ratios of the local
concentrations (in cm$^{-3}$) are taken to be equal to the ratios of
the column densities integrated over depth (in cm$^{-2}$).  This
assumption is accurate for molecules whose abundances peak in the
center of the cloud, but not necessarily for species such as radicals
whose abundances peak in the outer part of the envelope (see Figure 5
of van Dishoeck 1998a).

\subsubsection{The C$\,^+$ $\to$ C $\to$ CO transition}

The principal chemical characteristics of the depth- and
time-dependent models are governed by the transition of carbon from
atomic to molecular form.  In the depth-dependent case, C$^+$
recombines to C around $A_V\approx 1$ mag, followed by the conversion
to CO around $A_V\approx 2$ mag.  The CO photodissociation rate as a
function of depth is crucial in this transition (e.g., van Dishoeck \&
Black 1988, Lee et al.\ 1996b). The chemistry of other species follows
the C$^+$ $\to$ CO transition. At the edge, only the simplest diatomic
molecules are found. Around $A_V=1$--2 mag, the presence of both C and
C$^+$ and simple hydrides results in an increase in the abundance of
hydrocarbon molecules such as CN and C$_2$H. Deeper inside the cloud,
atomic carbon is no longer available, and destruction by O becomes
more important than photodissociation. Stable species such as CH$_4$,
C$_2$H$_2$ and HCN become dominant.

\begin{figure}[tb]
\vspace{-3.3cm}
\begin{center}
\leavevmode
\psfig{figure=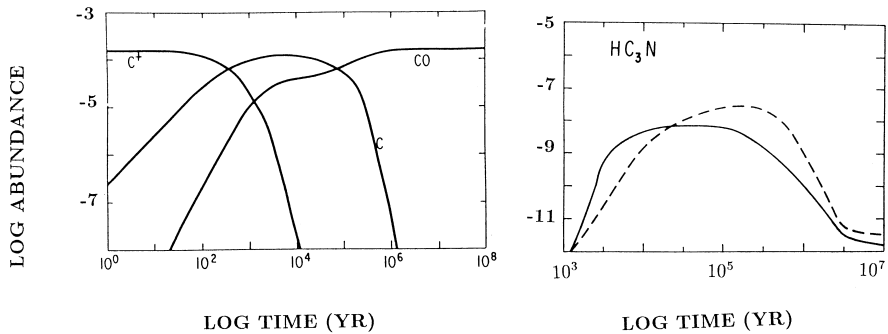,width=18cm,angle=90}
\end{center}
\vspace{-4.3cm}
\caption{
Time dependence of the abundances of C$^+$, C and CO (left)
and HC$_3$N (right) for a cloud with $n_{\rm H}=2\times 10^4$
cm$^{-3}$. The latter behavior is typical of all complex
carbon--bearing molecules
(from: Herbst \& Leung 1986). 
}
\label{fig-4}
\end{figure}

Many of the same features are observed in the pseudo time-dependent
models, if depth is replaced by time.  These models usually start
with the assumption that dark clouds originate from diffuse gas, so that
all species except H$_2$ are initially in atomic form, with carbon
present as C$^+$.  On a time scale of $\sim 10^3$ yr, C$^+$ recombines
to C, which subsequently transforms to CO after $\sim 10^5$ yr.  Since
the presence of atomic carbon is essential to build up more complex
organic species such as HC$_3$N, they are abundant only at early
times, but not at steady-state (see Figure~\ref{fig-4}).

The main time-dependent aspects of the other elemental pools ---O, N
and S---, depend on how their chemistry is linked to that of
carbon. For example, the abundance of CS has a different time behavior
than that of SO, because CS is formed by reactions with
carbon-hydrides, whereas SO is destroyed by atomic carbon.  This leads
to two classes of molecules: (i) species like CN, HCN, CS and complex
carbon chains, whose production is linked to the carbon network and
which have larger abundances at early times; and (ii) molecules such
as N$_2$, NH$_3$, N$_2$H$^+$ and SO that are independent or
destructively linked to the carbon chemistry and which exhibit higher
abundances at equilibrium.   

\subsection{Fractionation}

At the low temperatures of pre-stellar cores and the outer envelopes
of YSOs, significant enhancement of minor isotopes can occur as a
result of so-called `fractionation'. The most relevant case for
star-forming regions is the deuterium fractionation, initiated by the
isotope exchange reactions H$_3^+$ + HD $\to$ H$_2$D$^+$ + H$_2$ and
CH$_3^+$ + HD $\to$ CH$_2$D$^+$ + H$_2$ (Wootten 1987, Millar et al.\
1989). These reactions occur preferentially in the forward direction
at low temperatures, because the heavier H$_2$D$^+$ and CH$_2$D$^+$
are slightly more stable than H$_3^+$ and CH$_3^+$ due to their lower
zero-point vibrational energy. H$_2$D$^+$ and CH$_2$D$^+$ subsequently
transfer the deuterium to, for example, CO or N to form DCO$^+$ or
DCNH$^+$. The latter ion leads to DCN through dissociative
recombination.  The observed DCO$^+$/HCO$^+$ and DCN/HCN abundance
ratios are factors of 100 to $>$1000 larger than the overall [D]/[H]
ratio of $\sim 1.6\times 10^{-5}$.

\subsection{Successes and problems}

Pure gas-phase chemistry models have been remarkably successful in
explaining many observed features of interstellar chemistry (Herbst
1995, Turner et al.\ 1998). These include the abundances of simple
species in diffuse and translucent clouds; the presence of H$_3^+$ and
the related high abundances of protonated species such as HCO$^+$ and
N$_2$H$^+$ in dark clouds; the high abundances of unsaturated and
metastable molecules in dark clouds such as $l-$ and $c-$C$_3$H or
HNC; and the large isotopic fractionation.

Most of the problems with gas-phase models stem from uncertainties in
the basic reaction rates. For example, if reactions of bare carbon
chains C$_n$ with O or N are rapid, the observed abundances of
carbon-chain molecules like HC$_7$N cannot be reproduced, even at
early times.  Another complication is that time-dependent models have
more than one solution in certain regions of parameter space (e.g., le
Bourlot et al.\ 1995, Lee et al.\ 1998).  It is not yet clear, however,
what the astrophysical consequences of this `bistability' phenomenon
are. Finally, gas-phase models of dark clouds predict that a
substantial fraction of the oxygen is driven into O$_2$ at low
temperatures. Recent limits on the O$_2$/CO ratio of less than 0.03
are difficult to accomodate in such models (Olofsson et al.\ 1998,
Marechal et al.\ 1997, Melnick et al.\ 1998).

\section{Grain-surface chemistry}

\subsection{Basic surface processes}

The chemistry on the surfaces of interstellar grains has received
ample discussion in the literature (e.g., Tielens \& Allamandola 1987,
Herbst 1993, Tielens \& Charnley 1997, Langer et al.\ 1999).  Four
different steps can be distinguished: (i) accretion; (ii) diffusion;
(iii) reaction; and (iv) ejection or evaporation. The time scale for a
molecule to collide with a grain and accrete is given by $ t{_{\rm
acc}}\ \approx 2\times 10^9/n_{\rm H} y_S \ {\rm yr} $ where $y_S$ is
the sticking probability which is thought to lie between 0.1 and 1.0
for most species (see Williams 1993 for a review).  Thus, in envelopes
around YSOs with densities of order $10^5$ cm$^{-3}$ or larger, the
time scale for depletion is shorter than the collapse time (Walmsley
1991).  The timescales for diffusion and evaporation increase
exponentially with the binding energies of the species to the grain,
which in turn scale with its mass in the case of physical adsorption.
Under most interstellar circumstances, the rate-limiting step is the
accretion of new species rather than the diffusion over the surface.

\begin{figure}[tb]
\begin{center}
\leavevmode
\psfig{figure=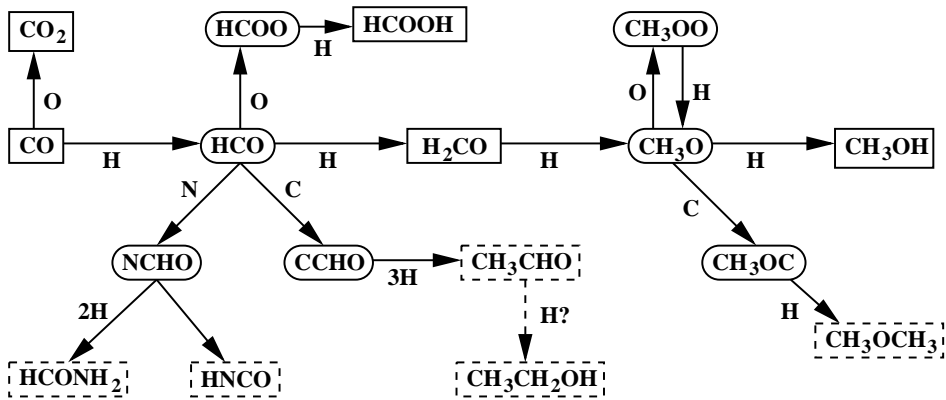,width=12.5cm,angle=-90}
\end{center}
\caption{Grain-surface chemistry routes involving CO. Solid rectangular boxes
contain molecules which have been identified in interstellar ices, whereas
dashed boxes indicate molecules detected in the gas phase in hot cores
(figure by Tielens, based on Tielens \& Hagen 1982).}
\label{fig-5}
\end{figure}

Because atomic hydrogen is so abundant and mobile, grain-surface
chemistry leads primarily to hydrogenated species such as H$_2$O,
NH$_3$ and CH$_4$.  Although many of the surface reactions have energy
barriers, they can still proceed because of the relatively long time
scale ($\sim$ 1 day) available for reaction before another species is
delivered from the gas to the grain.

The dominant molecule accreted from the gas is CO, and it therefore
forms a key in the surface chemistry. Hydrogenation of CO leads to
HCO, H$_2$CO and eventually CH$_3$OH (see
Figure~\ref{fig-5}). The intermediate HCO (formyl) and
CH$_3$OH (methoxy) radicals can also react with various other accreted
species, leading to more complex organic molecules such as CH$_3$CHO,
HCOOH and HCONH$_2$. The identification of many of these species in
molecular clouds, directly in ices or indirectly in hot cores
where the ices have evaporated off the grains, provides support for
this scheme.

Hydrogenation of molecules is particularly effective in the outer
regions of YSO envelopes where the atomic hydrogen abundance is high.
At densities greater than $\sim 10^5$ cm$^{-3}$ shielded from
ultraviolet radiation, atomic oxygen becomes more abundant than atomic
hydrogen, leading to oxidation of various accreted molecules. For
example, CO can react with O to form CO$_2$, and O$_2$ with O leads to
O$_3$.

Deuterium fractionation of molecules can also occur through
grain-surface chemistry, because the gas-phase atomic D concentration
inside dense clouds is enhanced compared with atomic H through
dissociative recombination of DCO$^+$ (Tielens 1983). The
atomic D can then react with molecules on the surface to form
deuterated species. Doubly-deuterated molecules such as D$_2$CO and
NHD$_2$, which have been detected in the envelopes around low- and
high-mass stars, are thought to be formed by this route (e.g.,
Ceccarelli et al.\ 1998).

\begin{table}[tbh]
\begin{center}
\caption{Interstellar ice composition and evaporation temperatures}
\label{tab-ice}
\begin{tabular}{lrrr}
\hline
Species & Abundance$^a$ & $T_{\rm ev}$$^b$ (K) \\
\hline
H$_2$O & 100 & 90 \\
CO     & 16  & 16 \\
CO$_2$ & 24  & 50 \\
CH$_4$ & 2   & 18 \\
CH$_3$OH & 4 & 80 \\
XCN      & 2 & ? \\
OCS      & $\ldots$ & 85 \\
NH$_3$   & 10$^c$ & 55 \\
N$_2$    & $\ldots$ & 13 \\
O$_2$    & $<30$$^d$  & 13 \\
\hline
\end{tabular}
\end{center}

$^a$ Abundances are relative to H$_2$O=100, and refer to NGC 7538 IRS9
(Whittet et al.\ 1996). Abundances are variable from source to source
due to differential outgassing and thermal processing. The H$_2$O
abundance is typically $10^{-4}$ with respect to H$_2$.\\

\vspace{-0.3cm}

$^b$ Evaporation temperatures for pure ices under
interstellar conditions; see text for mixed ices.\\  

\vspace{-0.3cm}

$^c$ Tentative identification by Lacy et al.\ (1998). \\

\vspace{-0.3cm}

$^d$ Vandenbussche et al.\ (1999).

\end{table}

\subsection{Thermal processing and evaporation}

In star-forming regions, the ice mantles formed in the cold
pre-stellar and collapse phases can be heated by the young star. This
leads to restructuring of the ice matrix and outgassing when the
temperature is near its sublimation point. Table~\ref{tab-ice}
summarizes the sublimation temperatures of various species under
interstellar conditions (Sandford \& Allamandola 1990, 1993); these
are typically lower than found in laboratory experiments because of
the lower pressures in space. Interstellar ices are not pure, but in
mixed ices, the evaporation of each component is largely determined by
its own sublimation behavior, unless the abundance of one of the
species is less than $\sim 5$\%.  Thus, an ice mixture of
H$_2$O/CO/CO$_2$=2/1/1 shows release of CO around 20~K, CO$_2$ around
50~K and H$_2$O around 90~K.  However, a mixture of
H$_2$O/CH$_4$=100/1 shows evaporation of both H$_2$O and CH$_4$ around
90~K.  Only a small fraction of the CO or CO$_2$ stays behind, and is
released when the whole H$_2$O ice rearranges from amorphous to
crystalline form.

Ices containing CH$_3$OH show particularly interesting behavior upon
heating, because CH$_3$OH has a sublimation temperature above that of
the phase transformation of H$_2$O. Thus, H$_2$O/CH$_3$OH ices heated
to $\sim 80$--90~K will segregate into rather pure H$_2$O- and
CH$_3$OH domains (Blake et al.\ 1991). Recent experiments indicate
that CO$_2$/CH$_3$OH mixtures show similar segregation behavior
(Ehrenfreund et al.\ 1998), and evidence for this process is seen in
the observed profiles of interstellar solid CO$_2$ toward massive YSOs
(Boogert et al.\ 1999, Gerakines et al.\ 1999).

\subsection{Non-thermal desorption}

Thermal evaporation is only efficient at dust temperatures $T_{\rm
dust}$ higher than 20~K.  Some desorption can also occur by
non-thermal processes in the cold pre-stellar cores and outer
envelopes with $T_{\rm dust}\approx 10$~K.  These processes have been
reviewed most recently by Schutte (1996), and include cosmic ray spot
heating, heating due to the energy liberated by the formation of
molecules on grains, heating due to grain--grain collisions, and
explosive heating due to exothermic reactions between stored radicals
on grains.  The desorption time scales depend strongly on the binding
energies of the molecules at the surfaces and the surfaces involved
(silicates, carbonaceous material, ices). All of them are particularly
effective for non-polar ices containing CO, O$_2$ and N$_2$, and less
for H$_2$O-rich ices which contain strong hydrogen bonds.

\subsection{Energetic processing}

The cosmic-ray induced ultraviolet photons and the radiation produced
by the young star can further process the ices. Ultraviolet photolysis
of H$_2$O, CO, NH$_3$ and CH$_4$ produces radicals which can react
with each other and with the parent molecules to form more complex
molecules like H$_2$CO, HCOOH and HCOCH$_3$ (e.g., Bernstein et al.\
1995, Gerakines et al.\ 1996). Much of the outcome of this chemistry
is very similar to that of the hydrogenation of CO, and further
quantitative studies are needed to assess their relative importance.

Photolysis of ice mixtures containing H$_2$O, CO, NH$_3$ and CH$_3$OH
often leaves a non-volatile organic residue containing a variety of
complex molecules (e.g., Schutte et al.\ 1993, Bernstein et al.\
1995).  Some of these chemical effects are also found in experiments
in which ices are bombarded with highly-energetic charged particles,
analogous to cosmic rays or X-rays (e.g., Moore et al.\ 1983, Kaiser
\& Roessler 1997, Teixeira et al.\ 1998).  The photochemical and
high-energy bombardment processes are often referred to as
`energetic' processing in the literature, without discrimination.

\subsection{Polar and apolar ices}

The ices in interstellar clouds are generally not homogeneous, but
consist of different layers or domains (e.g., Tielens et al.\ 1991,
Ehrenfreund et al.\ 1998). Two broad classes of ices can be
distinguished.  Polar ices are dominated by H$_2$O ice and contain
minor amounts of CO$_2$, CO, CH$_4$ and CH$_3$OH. Apolar ices are
dominated by species such as CO and perhaps some O$_2$ and N$_2$, but
contain very little H$_2$O. The two phases can be distinguished by
their line profiles, since the force constants of CO molecules
embedded in an H$_2$O-matrix will be slightly different from those of
CO surrounded by other CO molecules.

The different ice phases can arise from a combination of at least two
processes, caused by the density and temperature gradient in YSO
envelopes (see Schutte 1999 for review). First, the density gradient
results in a steep gradient in the gas-phase H/CO ratio. At low
densities or at the edge of the cloud, the high concentration of H
results in polar ices. Deep inside the envelope at high densities,
mostly apolar ices of accreted CO, O$_2$ and N$_2$ form. Layered ices
can be produced in a collapsing envelope with the polar ices
condensing first and the non-polar species forming a volatile
`crust'.  Second, the evaporation processes can shape the
composition of the ice mantles, because the desorption mechanisms are
much more efficient for volatile non-polar species like CO than for
non-volatile material like H$_2$O and CH$_3$OH. This `distillation'
effect decreases the apolar ices compared with the polar ices at
higher temperatures.

\subsection{Gas--grain models}

Based on the above processes, two different chemical regimes can be
distinguished in models which take both gas-phase and grain-surface
processes into account.  In the `accretion-limited' regime, the
diffusion time is much shorter than the accretion time so that a
species can diffuse on the surface until it finds a co-reactant. The
chemistry is limited by rate at which new species are delivered to the
surface.  In the `reaction-limited' regime, the opposite holds so that
many reactive species are present on a grain surface and the reaction
is controlled by surface concentrations as well as kinetic parameters.
Most of the gas--grain chemical models have been formulated in the
`reaction-limited' regime using rate equations which mirror those used
for gas-phase chemistry (e.g., Hasegawa \& Herbst 1993). This approach
is only accurate when a large number of reactive species exist on a
single grain surface, since only average abundances are
calculated. This condition is usually not met in interstellar clouds,
since the accretion times are long, grains are small, and reactions
are fast, so that at most one reactive species is present on a grain
at any time. The surface chemistry is therefore in the
accretion-limited regime and can only be properly treated by a
Monte-Carlo method which determines the likelihood of two such species
arriving from the gas in succession onto a particular grain in a
steady-state model (e.g., Tielens \& Hagen 1982). Recently, ad-hoc
modifications of the rate equations have been proposed to correct the
shortcomings of the reaction-limited approach (Caselli et al.\ 1998,
Shalabiea et al.\ 1998).

\subsection{Successes and problems}

The observations of H$_2$ in diffuse clouds and of large amounts of
ices in dense clouds can only be explained in models in which these
molecules are formed on the surfaces of interstellar grains and/or
accreted from the gas. Thus, gas--grain interactions play an essential
role in interstellar chemistry. The main problem is an accurate,
quantitative description. Many basic grain-surface reactions with H or
O at low temperatures are still poorly understood theoretically or
ill-constrained by laboratory experiments, partly because the details
of the surfaces of interstellar grains are still poorly characterized.

\section{Determination of molecular abundances}

The comparison of observations and chemical model calculations
ultimately depends on the translation of spectral-line data into
molecular abundances.  In the following, the observational techniques
will be discussed, and the methods for constraining the physical
structure and subsequently the chemical abundances will be outlined.

\subsection{Observational techniques}

The energy level structure of a molecule can be decomposed into an
electronic part $E_{\rm e\ell}$, a component due to the vibrational
motion of the nuclei $E_{\rm vib}$, and a component describing the
overall rotation of the molecule in space $E_{\rm rot}$. In general,
$E_{\rm e\ell} >> E_{\rm vib} >> E_{\rm rot}$: transitions
between two electronic levels lie typically in the visible/ultraviolet
part of the spectrum, those between two vibrational levels in the
infrared part, and those between two rotational levels at
(sub-)millimeter wavelengths. Because of the high extinction, only
infrared and millimeter observations are feasible in star-forming
regions.

\begin{figure}[tb]
\begin{center}
\leavevmode
\psfig{figure=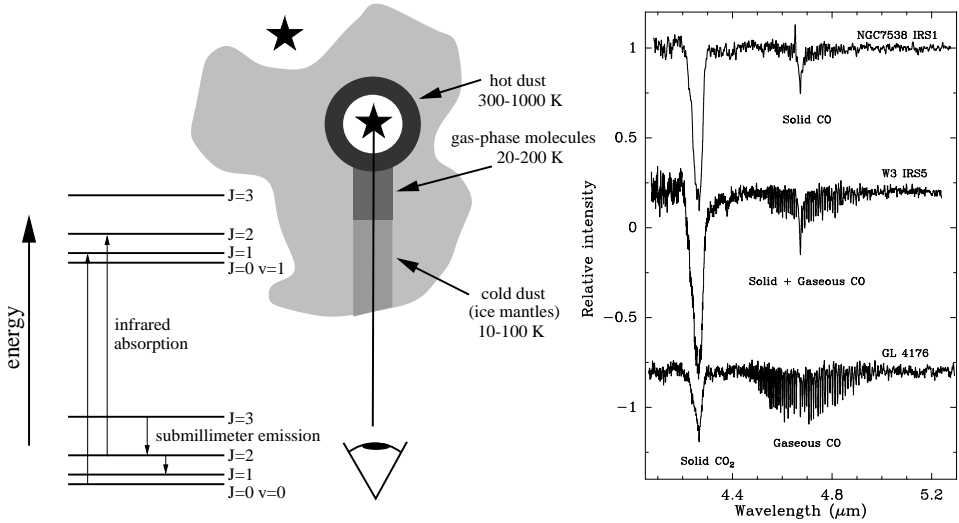,width=13cm,angle=-90}
\end{center}
\vspace{-0.5cm}
\caption{Left: Schematic illustration of infrared absorption line
observations of gas and dust toward embedded or background
sources. The infrared continuum is provided by the hot dust at
300--1000~K. Right: Normalized ISO--SWS spectra toward three massive
young stellar objects embedded in dense molecular clouds. The strong,
broad absorption at 4.27 $\mu$m is due to solid CO$_2$, whereas the
characteristic ro-vibrational P- and R-branch structure at 4.4--4.9
$\mu$m indicates the presence of warm, gaseous CO along the line of
sight (van Dishoeck et al.\ 1998).}
\label{fig-6}
\end{figure}

\subsubsection{Rotational line emission at (sub-)millimeter
wavelengths}

Most molecules listed in Table~\ref{tab-mol} are observed by their
rotational transitions within the ground electronic and vibrational
state (Figure~\ref{fig-6}).  The sensitivity of large single-dish
(sub-)millimeter telescopes allows the detection of weak lines and
rare chemical species and isotopes down to abundances of $10^{-11}$
with respect to hydrogen. However, not all molecules can be observed
by this method.  Symmetric molecules lack the permanent dipole moment
required for rotational emission lines, so that important species like
H$_2$, N$_2$, CH$_4$ and C$_2$H$_2$ cannot be observed in this way. In
addition, the Earth's atmosphere prevents ground-based observations of
low-lying transitions of O$_2$, CO$_2$ and H$_2$O.  Because of its
chemical stability, CO often serves as a tracer of H$_2$, assuming
either a fiducial CO/H$_2$ abundance ratio or a conversion factor
between integrated CO intensity and H$_2$ column density (see van
Dishoeck \& Black 1987, van Dishoeck 1998a for reviews).  The only
direct measurement of CO/H$_2$=$2.7\times 10^{-4}$ in the warm dense
cloud toward NGC 2024~IRS2 (Lacy et al.\ 1994) is a factor of 3 higher
than the value of $8\times 10^{-5}$ commonly used in dark clouds
(Frerking et al.\ 1982).  Submillimeter continuum emission from dust
can also act as a standard when values are adopted for the dust
emissivity, its temperature, and the dust-to-gas mass ratio.

Clouds can be mapped at angular resolutions between $10''$ (1500~AU at
150 pc) and a few arcminutes with single-dish instruments. Even in the
nearest star-forming clouds, regions of different physical or chemical
characteristics (envelope, outflow, disk, ...) are likely to fall
within one beam, complicating the derivation of molecular
abundances. Much higher angular resolution of the order of one
arcsecond is offered by interferometers. However, the sensitivity of
these instruments is lower than that of single-dish telescopes, and
their interpretation is complicated because they filter out all
emission on extended scales defined by the shortest baselines (cf.\ \S
8.1).

A major advantage of heterodyne (sub-)millimeter spectroscopy is the
very high spectral resolution, typically better than a few tenths of a
km~s$^{-1}$ and covering the full extent of the lines up to a few
hundred km s$^{-1}$. This allows separation of regions with different
physical conditions within one beam if they have different velocities
or line profiles. A drawback of rotational emission spectroscopy is
that the frequency separation between different transitions of the
same molecule is usually too large to be obtained in a single setting,
and multiple observations are needed to constrain the excitation,
often involving different telescopes or beam sizes.

\subsubsection{Vibrational absorption at infrared wavelengths}

Absorption of infrared emission into (ro-)vibrationally excited levels
of molecules offers a complementary view of the chemical content (see
Figure~\ref{fig-6}). Background stars or embedded objects of
sufficient infrared brightness can usually be found in star-forming
regions, often the YSO under study itself. Absorption lines probe
molecular gas along a pencil beam to the infrared source with very
high angular resolution, but mapping is not possible and different
conditions still exist along the line of sight. The velocity
resolution is generally lower, so that the line profiles are often not
resolved.

Because of the low spectral resolution and the intrinsic strengths of
the transitions, the abundance limit of infrared observations is two
to four orders of magnitude higher than that at millimeter
wavelengths. On the other hand, symmetric species like C$_2$H$_2$ and
CH$_4$ can be observed, because the vibrational modes induce a
temporary dipole moment. Different transitions of a molecule often
fall in a single infrared spectrum, providing direct constraints on
the molecular excitation up to high energy levels. By their very
nature, absorption lines also directly measure the line opacity, if
the profiles are resolved. Another major advantage is that infrared
studies are not limited to the gas phase. Molecules frozen onto dust
grains show a markedly different line profile, since the rotational
degrees of freedom are collapsed into a single, broad feature which is
slightly shifted from the position of that of the gas-phase molecule
(Figure~\ref{fig-6}).  This offers the possibility to study the
relative amounts of a species present in the gas-phase and in the
condensed phase, and to probe the ice mantles as formation sites of
various molecules like CH$_3$OH and H$_2$CO.
The {\it Short Wavelength Spectrometer} (SWS) on ISO is particularly
well suited for such studies (see \S 6.2, Figure~\ref{fig-6}
and \ref{fig-10}).

\subsection{Constraining the physical structure}

In order derive molecular column densities or abundances from the
observations, a good physical model of the region is a prerequisite.
Observations of different rotational lines provide very useful probes,
because they are collisionally excited by H$_2$ for densities of
$10^2$--$10^7$ cm$^{-3}$ and temperatures of 5--200 K (Evans 1980,
Walmsley 1987, Genzel 1992).

The statistical equilibrium equations for the level populations $n_i$
due to radiative and collisional processes are given by
$$ n_i \sum_{j<i}A_{ij} + n_i \sum_j(B_{ij} J + C_{ij}n_{\rm col}) = 
   \sum_{j>i} n_j A_{ji} + \sum_j n_j (B_{ji}J + C_{ji}n_{\rm col}), $$
where the left-hand side represents the processes leading to the loss
of population of level $i$ and the right-hand side the gain of
population.  Here $A_{ij}$ and $B_{ij}$ are the Einstein $A$ and $B$
coefficients for spontaneous and stimulated emission and absorption,
$J$ is the intensity of the radiation field averaged over all
directions and integrated over the line profile, $C_{ij}$ is the
collisional rate coefficient between levels $i$ and $j$, and $n_{\rm
col}$ is the density of collision partners, primarily H$_2$.  The
upward and downward rate coefficients are related through detailed
balance by $ C_{ji} = (g_i/g_j) C_{ij} \exp({-h\nu/kT}). $ The values
of $C_{ij}$ for collisions of different molecules with H$_2$ $J$=0
have largely been derived from detailed quantum mechanical
calculations (Green 1975, Flower 1990), and form one of the largest
uncertainty in analyzing molecular excitation.  Information on
collisional rate coefficients at high temperatures $>$200~K is still
scarce, and collisions with H$_2$ $J$=1 or higher are generally not
taken into account explicitly.

\begin{figure}[tb]
\vspace{-3.0cm}
\begin{center}
\leavevmode
\psfig{figure=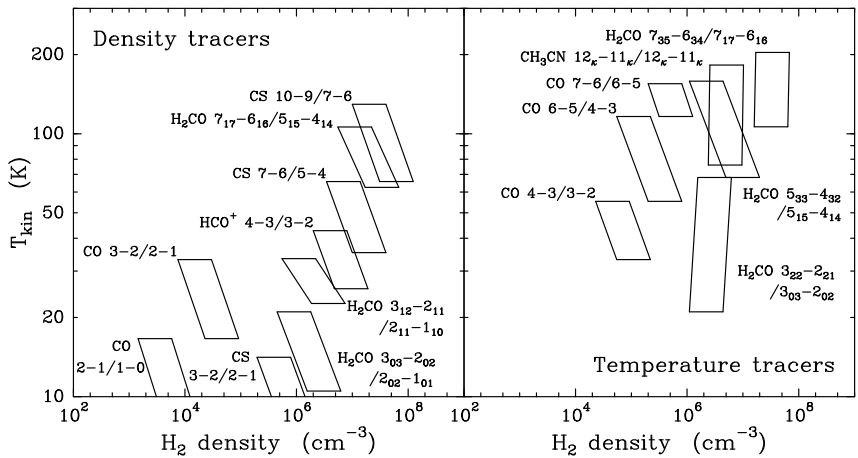,width=15.5cm,angle=-90}
\end{center}
\vspace{-1.3cm}
\caption{Examples of molecular line ratios which are  used to 
constrain the density and temperature structure of YSO envelopes.
}
\label{fig-7}
\end{figure}

The critical density for a given transition is the density at which
the rates for collisional processes become comparable to those for
radiative processes, resulting in substantial population of level
$i$. In formula, $ n_{\rm crit}^i = A_{ij} / \sum_j C_{ij}. $ If the
transition becomes optically thick, the critical density is lowered by
$1/\tau$, since part of the emitted photons are re-absorbed. Here
$\tau$ is the optical depth of the line. Note that higher frequency
transitions have higher critical densities, since $A_{ij} \propto
\nu^3$. Also, molecules with large dipole moments $\mu$ have high
critical densities, since $A_{ij} \propto \mu^2$. Thus, by choosing
the appropriate molecule and transition, a large range of densities
can be probed. This is schematically indicated in
Figure~\ref{fig-7}, which shows the range of physical parameters
for which different line ratios are most sensitive.

The contribution of the line photons to the excitation is taken into
account by a variety of formalisms, including the large velocity
gradient, Sobolev, microturbulent, or escape probability methods
(e.g., Sobolev 1960, Habing 1988, Osterbrock 1989 Appendix 2).

\subsection{From observations to abundances}

\begin{figure}[tb]
\begin{center}
\leavevmode
\psfig{figure=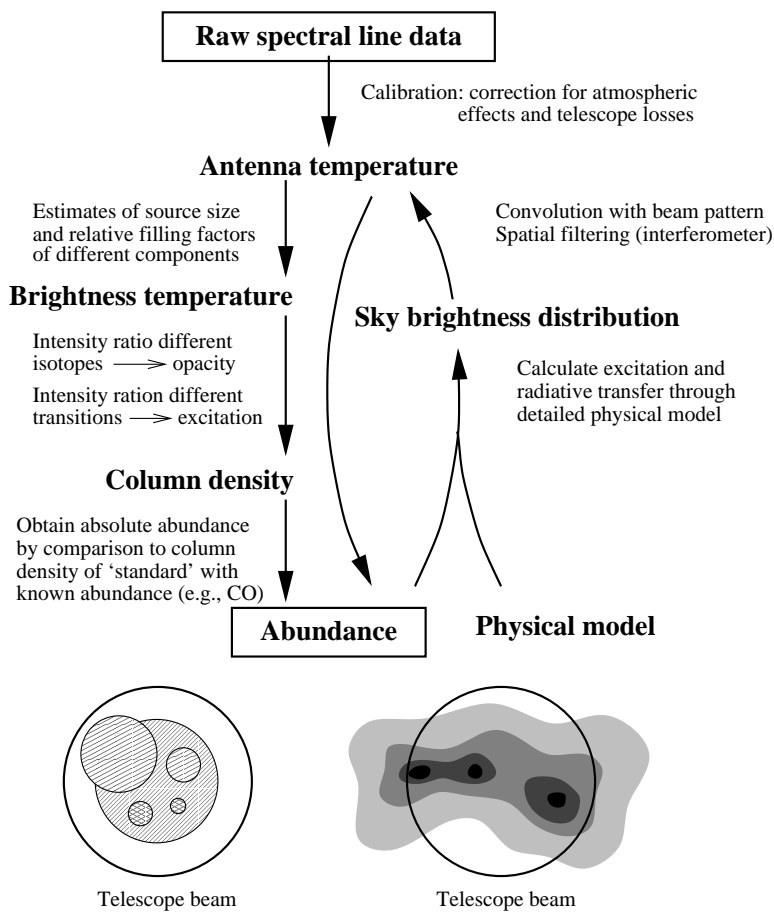,height=12cm,angle=270}
\end{center}
\caption{The steps involved in deriving molecular abundances from
spectral line data. The left-hand side shows the `homogeneous'
approach, the right-hand side the `detailed' approach.
}
\label{fig-8}
\end{figure}

Figure~\ref{fig-8} gives a schematic overview of the steps involved
in the determination of abundances from observations.  Correcting for
atmospheric effects and telescope losses yields the antenna
temperature. This is the actual sky brightness averaged over the main
forward beam of the telescope, or, for interferometric observations,
the sky brightness averaged over the synthesized beam size after
spatial filtering by the array baselines.

Relating the antenna temperature to the underlying brightness
distribution on the sky requires assumptions about the source
structure. Figure~\ref{fig-8} illustrates two extreme approaches to
this problem. The `homogeneous' method, shown on the left, uses a
single set of physical conditions and a given size, or beam-filling
factor, for the source, which transforms the beam-averaged antenna
temperature into brightness temperature on the sky. The molecular
excitation and opacity relate this brightness temperature to the
column density of the molecule. The line opacity follows from the
intensity ratio of lines of different isotopic varieties. The
excitation can either be assumed to be in local thermodynamic
equilibrium, valid if the density exceeds the critical density for the
observed transition, or can be derived from statistical equilibrium
and radiative transfer calculations as described above once the
physical parameters have been constrained.  Repeating the procedure
for an appropriate abundance standard like CO yields the molecular
abundance relative to H$_2$.

More complex source structures can be described with multiple
`homogeneous' components, each with its own beam-filling factor. This
approach works well if the number of different components is limited,
for example, if the cloud is described as a collection of dense clumps
embedded in a more diffuse interclump medium (e.g., Hogerheijde et
al.\ 1995). When the beam is filled with a large number of different
physical regimes, as is the case, e.g., for the power-law density and
temperature distributions often used to describe YSO envelopes, the
multi-component `homogeneous' approach no longer suffices. The
right-hand side of Figure~\ref{fig-8} shows a more `detailed'
method. Starting with a fiducial source model, the molecular
excitation and the radiative transfer throughout the cloud is
calculated. Monte-Carlo techniques are often used in this step (e.g.,
Bernes 1979). The emergent sky brightness distribution is convolved
with the telescope beam, or sampled at the interferometer's spatial
frequencies. Matching the observed and model antenna temperature
constrains the molecular abundance.

The interpretation of interferometric observations, and the derivation
of molecular abundances from them, requires that spatial filtering by
the interferometer is taken into account explicitly. Even if the data
sets of the molecule in question and that of a standard like C$^{18}$O
contain the same spatial frequencies, both lines can have very
different distributions of intensity over those frequencies if they
trace different physical regimes. Different levels of flux are
resolved out by the interferometer, resulting in abundance estimates
deviating by factors of a few or more if not accounted for properly.

\section{Chemistry in pre-stellar cores}

In the following sections, an overview will be given of recent
observations and models of star-forming regions at different
evolutionary stages, using the material presented in \S 1-4 as general
background information.  The discussion focuses on those species whose
abundances are most affected at a particular evolutionary state.
These molecules are often only minor components in terms of the
overall composition of the gas or dust.  Recent overviews of the major
reservoirs of the principal elements C, N and O are given, e.g., in
van Dishoeck et al.\ (1993) and van Dishoeck \& Blake (1998).  In
brief, most of the oxygen is contained in solid silicates, oxides and
ices, as well as gas-phase O, CO and, in warm regions, H$_2$O. Most of
the carbon is in some solid carbonaceous form, with the remainder in
gas-phase CO. Nitrogen is mostly locked up in gas-phase N$_2$ and N,
but the amount in solids is poorly constrained.

\subsection{Translucent clouds}

In order to study the effect of YSOs on the chemistry and test the
basic chemical networks, the abundances in quiescent clouds prior to
star formation need to be constrained.  The best clouds for this
purpose are the translucent and high-latitude clouds with visual
extinctions of a few mag and densities of a few thousand
cm$^{-3}$. These clouds have been studied by optical absorption lines
against reddened stars (e.g., Gredel et al.\ 1993), by millimeter
absorption lines against distant radio sources (e.g., Lucas \& Liszt
1997) and by millimeter emission observations (e.g., Turner 1996, 1998
and references cited). No signs of star formation have been found in
these regions.  The chemistry is characterized by simple diatomic and
triatomic molecules, radicals and ions.  Detailed models by Turner et
al.\ (1998) show that the observed abundances of most species are well
reproduced by the ion-molecule networks. The main exceptions are
formed by NH$_3$, H$_2$CO and H$_2$S, whose formation is probably
dominated by grain-surface chemistry.

\subsection{Dark cloud cores}

More than 100 dark cores have been identified through optical
extinction and molecular line studies, at least half of which have not
yet formed any stars (Myers \& Benson 1983, see chapter by Myers).
Most chemical studies have been directed toward TMC-1, a small
elongated, dense ridge of $\sim 0.6 \times 0.06$ pc in Taurus with a
mass of less than 40 M$_{\odot}$.  This clump shows a particularly
rich chemistry, with a large chemical gradient across the core. NH$_3$
and other `late' molecules peak in the northern part, whereas long
carbon-chain molecules such as HC$_7$N and C$_4$H are most abundant
in the southern part (Olano et al.\ 1988, Pratap et al.\ 1997).  High
spectral and spatial resolution observations show that there are at
least three different velocity components in this region on less than
10,000 AU scales, with different intensity ratios among the molecules
(Langer et al.\ 1997).  The abundances of the carbon-chains range from
$10^{-11}$ up to $10^{-8}$ and are several orders of magnitude larger
than found in other dark clouds. Such large abundances are usually
interpreted with pseudo time-dependent models at `early times'
$t\approx 10^5$ yr, assuming that the cloud evolved from a diffuse
cloud phase in which carbon was initially in atomic form (e.g., Lee et
al.\ 1996a).

\begin{figure}[tb]
\vspace{-1.7cm}
\begin{center}
\leavevmode
\psfig{figure=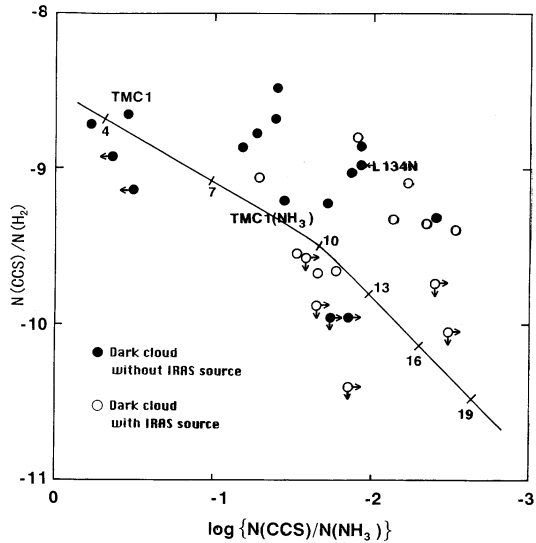,height=12.8cm,angle=90}
\end{center}
\vspace{-3.8cm}
\caption{Observed CCS abundance versus the CCS/NH$_3$ abundance
ratio in dark clouds with and without
young stars. The solid line indicates the model abundance ratios
for $T$=10~K and $n$(H$_2$)=$10^4$ cm$^{-3}$. 
The numbers along the line indicate
the `age' of the cloud in $10^5$ yr, assuming that the gas is
initially atomic-carbon rich
(from: Suzuki et al.\ 1992).
}
\label{fig-9}
\end{figure}

In order to investigate whether TMC-1 is chemically peculiar, Suzuki
et al.\ (1992) and Benson et al.\ (1998) performed a systematic study
of a few characteristic molecules (C$_2$S, HC$_3$N, HC$_5$N,
N$_2$H$^+$ and NH$_3$) in a large set of dark cores.  The abundances
of the carbon chains are found to correlate well with each other, but
not with NH$_3$ or N$_2$H$^+$, consistent with the case of TMC-1.
The observed C$_2$S/NH$_3$ abundance has been reproduced
quantitatively in models which start from diffuse gas and form dense
cores over a period of $10^5$ to $2\times 10^6$ yr 
(see Figure~\ref{fig-9}). In this scenario,
TMC-1S is one of the youngest clouds.  There are numerous other
non-equilibrium processes, however, such as penetration of ultraviolet
radiation, outflows or turbulence, which can lead to the breakdown of
CO into C or C$^+$ and therefore give the cloud core a chemically
`young' appearance.  Thus, the `chemical age' does not necessarily
measure the true age of the core, but indicates the time since some
dynamical event produced an atomic-carbon rich phase.

The model results are also sensitive to the [C]$_g$/[O]$_g$ elemental
abundance ratio in the gas. If this value is increased from the
canonical ratio of 0.4 to higher values, improvement with observations
for the complex molecules is obtained (Bergin et al.\ 1997, Terzieva
\& Herbst 1998).  Pratap et al.\ (1997) show that the observed
chemical gradient in TMC-1 can be reproduced by changes in
[C]$_g$/[O]$_g$ along the ridge caused by a density gradient.

This point is further illustrated by high spatial resolution
observations of the L1498 pre-stellar core by Kuiper et al.\ (1996),
which show a chemically differentiated structure with NH$_3$ peaking
in the inner region and C$_2$S in the atomic-carbon rich (clumpy)
outer part. Further chemical studies of such cores, especially those
which have a more centrally concentrated density structure and appear
to be on the verge of collapse (e.g., Ward-Thompson et al.\ 1994,
Motte et al.\ 1998), are warranted.

The ice mantles in quiescent dark clouds can be probed through the
observation of field stars behind the clouds (e.g., Chiar et al.\
1995, Whittet et al.\ 1998). H$_2$O, CO and CO$_2$ ices have been
detected, which indicate that up to 40\% of the heavy elements may be
frozen out at densities of a few $\times 10^4$ cm$^{-3}$ (Schutte
1999).  Evidence for significant CO depletion deep inside dark clouds
($A_V>20$~mag) also comes from comparison of C$^{18}$O observations
with extinction maps derived from infrared star counts 
(Lada et al.\ 1994, Kramer et al.\ 1999).

\subsection{Ionization fraction}

The ionization fraction $x$(e)=$n$(e)/$n$(H$_2$) is an important
parameter in the dynamical evolution of star-forming regions (see
chapters by McKee, Shu, and Mouschovias). In model calculations, $x$(e)
drops from its high value of $\sim 10^{-4}$ at the edge of the core to
$\sim 10^{-8}$ in the center, and scales roughly with $(\zeta/n_{\rm
H})^{1/2}$ (Millar 1990, Bergin \& Langer 1997).  At the edge, C$^+$
is the main supplier of electrons, whereas deep inside metal ions such
as Mg$^+$ and Fe$^+$, as well as molecular ions such as H$_3^+$,
HCO$^+$ and H$_3$O$^+$ dominate.

Recent observations of DCO$^+$/HCO$^+$ in a sample of dark cores give
ionization fractions of a few $\times 10^{-7}$, for an adopted cosmic
ray ionization rate of $5\times 10^{-17}$ s$^{-1}$ (Plume et al.\
1998, Williams et al.\ 1998).  Surprisingly, no systematic trends
between pre-stellar cores and cores with stars have been found, but
this may be due to the large scale of the observations ($\sim$10,000
AU).  A somewhat lower ionization fraction of $10^{-8}$--$10^{-7}$ is
found for more massive cores (Bergin et al.\ 1999, de Boisanger et
al.\ 1996).

\section{Chemistry in cold envelopes around YSOs}

\subsection{Models}

In the initial stages of collapse, the density increases strongly
whereas the temperature stays low, $T\approx 10$~K.  The principal
prediction of the models is enhanced freeze-out of molecules onto the
cold grains under these conditions.  The only exceptions are H$_2$,
He, H$_3^+$ and perhaps N$_2$. Models appropriate for the cold outer
envelopes have been made by Rawlings et al.\ (1992), Willacy et al.\
(1994), Bergin \& Langer (1997) and Shalabiea \& Greenberg (1995),
using parametrized fits to the density profiles in simple collapse
models such as that of Shu (1977) or Basu \& Mouschovias (1994).  They
differ strongly in the adopted mechanisms that return molecules to the
gas, ranging from none in Rawlings et al.'s to efficient desorption in
Shalabiea \& Greenberg's. The results are also sensitive to the
adopted binding energies on the icy mantles, especially whether the
outer layer is H$_2$O-rich or CO-rich.  The ions HCO$^+$ and
N$_2$H$^+$ are predicted to be good tracers of cold envelopes, because
their abundances remain high owing to the increase in the H$_3^+$
abundance when its main removal partners (CO, O, ...)  are depleted.
The use of HCO$^+$ to trace the envelope structure has been
demonstrated by Hogerheijde et al.\ (1997, 1998).

\subsection{Observations}

Observationally, depletion in collapsing cores is very difficult to
prove, because every cloud has a `skin' in which the abundances are
close to normal (Mundy \& McMullin 1997). Even if the skin amounts to
only a few \% of the total column density, its factor of 10--100
higher abundances can effectively mask any depletions deep inside.
The dust obscuration in the deeply embedded class 0 stage is still so
high that the young stars are too weak at near- and mid-infrared
wavelengths for direct observations of ices. Careful modeling of the
line and dust emission appears the only way to probe the abundances.
One of the best-studied cases is that of NGC 1333 IRAS4, where
depletions of more than a factor of 10 have been inferred (Blake et
al.\ 1995).  This phase of high depletion appears short-lived,
however, since for other class 0 objects like Serpens SMM~1 much less
freeze-out has been inferred (see \S 8.1).

\begin{figure}[tb]
\vspace{-0.9cm}
\begin{center}
\leavevmode
\psfig{figure=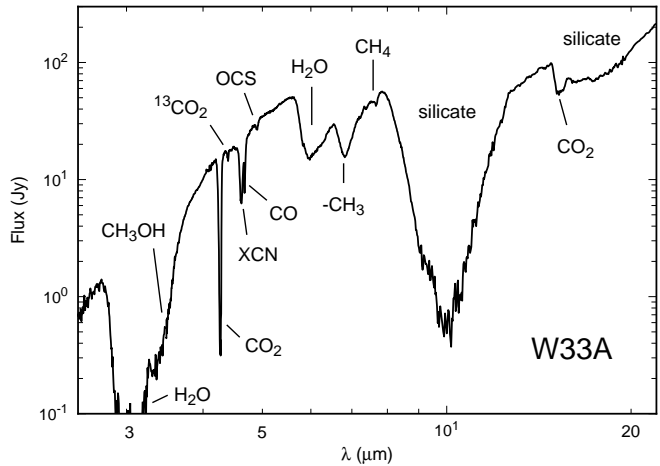,width=10.5cm,angle=90}
\end{center}
\vspace{-1.0cm}
\caption{ISO--SWS spectrum of the deeply embedded massive YSO W~33A.
Various absorption features due to silicate grain cores and icy mantles
are indicated (from: Gibb et al.\ 1999).}
\label{fig-10}
\end{figure}

At the more evolved, but still deeply embedded class~I stage, infrared
absorption line studies of the ice mantles in the cold envelopes
become possible, both for low- and high-mass objects (Whittet 1993,
Tielens \& Whittet 1997, Chiar et al.\ 1998, Teixeira et al.\
1998). Ground-based and ISO studies allow a nearly complete inventory
of these ices to be obtained for deeply embedded massive young stars
such as NGC 7538 IRS9, W33~A and RAFGL 7009S (Whittet et al.\ 1996,
d'Hendecourt et al.\ 1996, Ehrenfreund et al.\ 1997)
(see Figure~\ref{fig-10}).  The molecules that have been
identified and their typical abundances are summarized in
Table~\ref{tab-ice}. Although only abundances down to $\sim 0.5$\% of
H$_2$O-ice ($\sim 10^{-7}$ with respect to H$_2$) can be probed, the
inferred mantle composition is remarkably simple, consisting of
species resulting from the hydrogenation and oxidation of O, C, N and
CO.  The total column density of gas-phase CO is larger than that of
solid CO integrated along the line of sight for all sources studied so
far.

\section{Chemistry in warm envelopes around newly-formed stars}

When a young star begins to heat the inner surrounding envelope by
radiation and/or shocks above $\sim 100$~K, various distinct chemical
regions occur, including the hot core region, the region of
interaction of the outflow with the surrounding envelope, and the more
extended radiatively-heated warm envelope (see Figure~\ref{fig-11}).

\subsection{Warm envelopes}

\begin{figure}[tb]
\begin{center}
\leavevmode
\psfig{figure=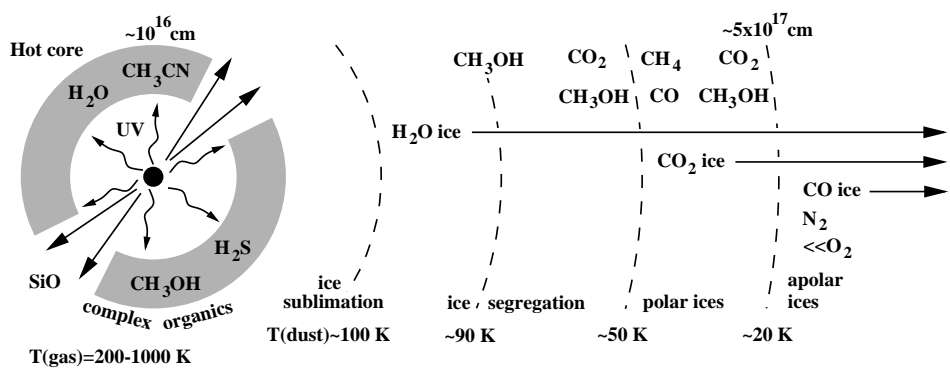,width=12.5cm,angle=270}
\end{center}
\caption{Schematic illustration of the physical and chemical
environment of massive YSOs. The variation in the structure of the ice
mantles due to heating and thermal desorption is shown (based on
Tielens et al.\ 1991, Williams 1993, van Dishoeck \& Blake 1998).  }
\label{fig-11}
\end{figure}

Self-consistent models of the thermal balance, chemistry and radiative
transfer in spherical envelopes heated by an internal source have been
constructed for low-mass YSO's by Ceccarelli et al.\ (1996) and for
high-mass YSOs by Doty \& Neufeld (1997).  Calculations of the radial
dust temperature distribution include those of Campbell et al.\
(1995), Kaufman et al.\ (1998), and van der Tak et al.\ (1999).  For a
$\sim 10^5$ L$_{\odot}$ source, the region where $T_{\rm dust}\approx T_{\rm
gas}>90$~K extends to $\sim 10^{16}$ cm, resulting in evaporation of
H$_2$O and CH$_3$OH ices and an enhancement of the gas-phase
abundances of these species by a factor of 100--1000 (see
Figure~\ref{fig-11}). The chemistry in this inner hot core is
discussed in \S 7.2. The region where $T>20$~K extends to $\sim
10^{17}$ cm.  An increase in the temperature from 10 to 100~K does not
have a significant effect on the gas phase chemistry, but does result
in evaporation of volatile species like CO.


Observational data on molecules in star-forming regions are scattered
throughout the literature, using a variety of (sub-)millimeter
telescopes with different beam sizes. Comprehensive single-dish
studies of a number of molecules have been performed for only a few
sources, including W~3 IRS5 (Helmich \& van Dishoeck 1997, see \S 8.2)
and NGC 2264 IRS1 (Schreyer et al.\ 1997). More detailed physical
models are needed to disentangle the contributions of changing
temperatures and densities in the envelope from possible radial
abundance gradients.

Important complementary information is obtained from high-resolution
infrared observations. Mitchell et al.\ (1990) showed the presence of
both cold ($T<60$~K) and hot ($T=120$--1000~K) gas along the lines of
sight toward massive YSO's from $^{12}$CO and $^{13}$CO
observations. The hot gas contains high abundances of C$_2$H$_2$, HCN
and CH$_4$ (Lacy et al.\ 1989, 1991; Evans et al.\ 1991; Carr et al.\
1995; Boogert et al.\ 1998; Lahuis \& van Dishoeck 1999), which are
enhanced by at least 1--2 orders of magnitude compared with the colder
envelope. Gas-phase CO$_2$ has a surprisingly low abundance (van
Dishoeck et al.\ 1996, Dartois et al.\ 1998,
van Dishoeck 1998b). Clear variations in the
gas/solid ratios are seen for various objects, indicating the
development of a hot core in the inner envelopes.

\subsection{Hot core regions}

Hot cores are small ($<$0.1 pc), dense ($n$(H$_2$)$> 10^6$ cm$^{-3}$)
and warm ($T\approx 200$~K or higher) regions located close to massive
young stars (see Walmsley \& Schilke 1993, Ohishi 1997 for
reviews). Observationally, hot cores are characterized by high
abundances of fully hydrogenated molecules such as NH$_3$, H$_2$S, and
H$_2$O (e.g., Gensheimer et al.\ 1996), as well as saturated complex
organic molecules like CH$_3$OH, CH$_3$CN, CH$_3$OCH$_3$ and
HCOOCH$_3$ (e.g., Hatchell et al.\ 1998). Examples of hot cores
include the Orion hot core and compact ridge (e.g., Blake et al.\
1987, Sutton et al.\ 1995), SgrB2(N) (e.g., Kuan \& Snyder 1996),
W~3(H$_2$O) (e.g., Helmich \& van Dishoeck 1997) and objects near
ultracompact H II regions such as G34.3+0.15 (Macdonald et al.\ 1996;
see chapter by Churchwell). The observed abundances of the species can
vary significantly from region to region.

\begin{figure}[tb]
\begin{center}
\leavevmode
\psfig{figure=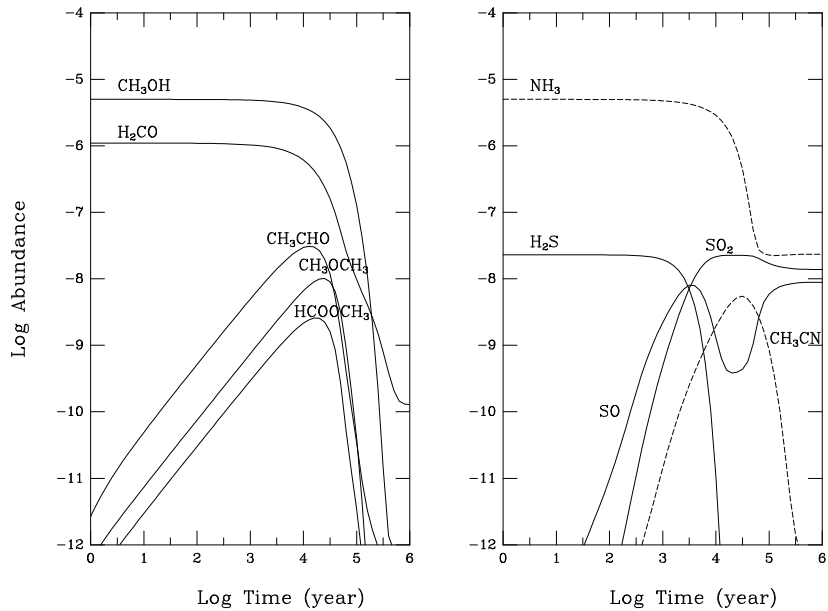,height=9cm,angle=270}
\end{center}
\vspace{-0.8cm}
\caption{Models of the gas-phase chemical evolution during the
`hot core' phase for $T$=200~K. At $t$=0, molecules such as CH$_3$OH,
NH$_3$ and H$_2$S are evaporated from the grain mantles into the gas
phase, where they drive a rapid chemistry leading to more complex
organic molecules such as CH$_3$OCH$_3$, HCOOCH$_3$ and CH$_3$CN.
The abundances peak after $10^4-10^5$ years (based on Charnley et al.\ 1992,
Charnley 1997).
}
\label{fig-12}
\end{figure}

The rich chemistry in hot cores is thought to be driven by the
evaporation of icy mantles due to heating near the young
star. Charnley et al.\ (1992) showed that the observed abundances in
the Orion hot cores can be reproduced if a mixture of simple ices
containing H$_2$O, CO, CH$_3$OH, NH$_3$ and/or HCN is evaporated into
the warm gas.  These molecules subsequently drive a rapid gas-phase
chemistry leading to complex organic molecules (e.g., Charnley et al.\
1992, Caselli et al.\ 1993, Charnley 1997).  The abundances of the
complex species peak after $\sim 10^4$ yr, thus offering a chemical
`clock' since the time of formation of the hot core (see
Figure~\ref{fig-12}).  At later times, the normal ion-molecule
chemistry takes over, resulting in destruction of the complex organic
molecules after $\sim 10^5$ yr.

Since methanol is one of the most abundant ice mantle consitutents, it
plays a crucial role in the hot core chemistry. Specifically, complex
oxygen-containing organics can be formed by protonation of CH$_3$OH by
H$_3^+$, H$_3$O$^+$ or HCO$^+$, followed by transfer of a -CH$_3$
group to a neutral molecule, e.g., CH$_3$OH$_2^+$ + CH$_3$OH $\to$
(CH$_3$)$_2$OH$^+$ + H$_2$O. Dissociative recombination of the ion then
produces di-methyl ether, CH$_3$OCH$_3$.  The detailed chemistry of
interstellar alcohols leading to large esters and ethers has been
discussed by Charnley et al.\ (1995).  Nitrogen-containing organics
can be formed through reactions of carbon ions with evaporated NH$_3$,
or through reactions of CN with evaporated C$_2$H$_2$.  In general,
the time scale for formation of nitrogen-rich molecules is somewhat
longer than that of oxygen-bearing species.

The H$_2$S/SO$_2$ ratio may be a particularly sensitive clock
(Charnley 1997, Hatchell et al.\ 1998).  Most of the sulfur is thought
to be in the form of H$_2$S in the ices, although this has not yet
been confirmed observationally.  Upon evaporation, H$_2$S is destroyed
by reactions with atomic H to atomic S at temperatures of a few
hundred K. S can subsequently react with OH and O$_2$ to form SO and
SO$_2$ on a timescale of $\sim 10^4$ yr.  The observed high D/H ratios
in hot core molecules (e.g., Millar \& Hatchell 1998) largely reflect
the high D/H ratios in the ices, which result from the cold dark cloud
phase prior to star formation.

\subsection{Outflow regions}

The earliest stages of star formation are accompanied by powerful
outflows, which originate within the young star/accretion disk
boundary region.  As the high velocity gas flows outward, it strikes
the envelope and creates shocks, which alter the chemistry.  Recent
reviews of the physical and chemical structure of shocks can be found
in Draine \& McKee (1993), Hollenbach (1997) and Bachiller (1997, this
volume).

If the shocks have high velocities ($>$50 km s$^{-1}$), the
temperature increases to such high values ($>10^4$~K) that most
molecules are collisionally dissociated. In addition, the ultraviolet
radiation from such $J$-type shocks can dissociate molecules both
within and ahead of the shock. The molecules reform slowly in a
lengthy, warm zone in the wake of the shock (Neufeld \& Dalgarno
1989a,b).

\begin{table}[tb]
\begin{center}
\caption{Physical regions of class~0 YSOs traced by various
observations} 
\label{tab-comp}
\begin{tabular}{ll}
\hline
Region & Observation \\
\hline
Bulk of the (cold) envelope & 
   Interferometer: Continuum on short spacings \\
 & Single-dish: Lines tracing $T_{\rm kin}< 40$ K \\
 & (e.g., C$^{18}$O 2--1; C$^{17}$O 3--2; H$^{(13)}$CO$^+$ 3--2, 4--3) \\
Warm, inner regions of the envelope &
    Interferometer: Continuum on long spacings \\
  & Single-dish: Lines tracing $\sim 100$ K \\
  & (e.g., $^{13}$CO 6--5; H$_2$CO $3_{22}$--$2_{21}$) \\
Outflow: gradual entrainment (?) &
   Interferometer: V or U shaped structures \\
 & (e.g., HCO$^+$, HCN) \\
 & Single-dish: line wings (e.g., CO, HCN) \\
Outflow: direct impact & 
   Single-dish and interferometer: Products of \\
 & ice mantle evaporation and grain destruction \\
 & (e.g., SiO, SO) \\
\hline
\end{tabular}
\end{center}
\end{table}

For lower shock speeds in clouds with low fractional ionization ($x_e
< 10^{-6}$), the shock is of $C$-type and the temperatures reach peak
values of only 2000--3000~K, insufficient to dissociate the
molecules. Reactions with energy barriers such as O + H$_2$ and S +
H$_2$ rapidly drive the available gas-phase oxygen and sulfur into
H$_2$O and H$_2$S.

In addition to these gas-phase processes, grain cores and mantles are
affected by shocks. The high-velocity $J$-shocks can cause destruction
or thermal sputtering of the cores, resulting locally in much enhanced
gas-phase abundances of Si, Fe, ... The lower velocity $C$-shocks can
inject a variety of refractory and volatile species into the gas phase
through non-thermal sputtering (e.g., Pineau des For\^ets \& Flower
1997). The released Si can then react with OH and O$_2$ to form SiO,
one of the principal diagnostics of shocks (e.g., Schilke et al.\
1997).

Outflows have been studied observationally in a variety of molecules,
including H$_2$ near-infrared emission, CO and SiO.  One of the best
studied regions chemically is L1157 (Bachiller \& Per\'ez Guti\'errez
1997). Both refractory (SiO, SO, ...) and volatile mantle species
(HCN, H$_2$CO, CH$_3$OH, ...) are found to be enhanced in the gas by
factors ranging from a few to $10^6$.  Strong H$_2$O lines have been
observed in outflow regions from the ground (e.g., Cernicharo et al.\
1994) and with the ISO satellite, indicating H$_2$O abundances of a
few times $10^{-5}$ (e.g., Liseau et al.\ 1996, Ceccarelli et al.\
1998, van Dishoeck et al.\ 1998).

\begin{figure}[tb]
\begin{center}
\leavevmode
\psfig{figure=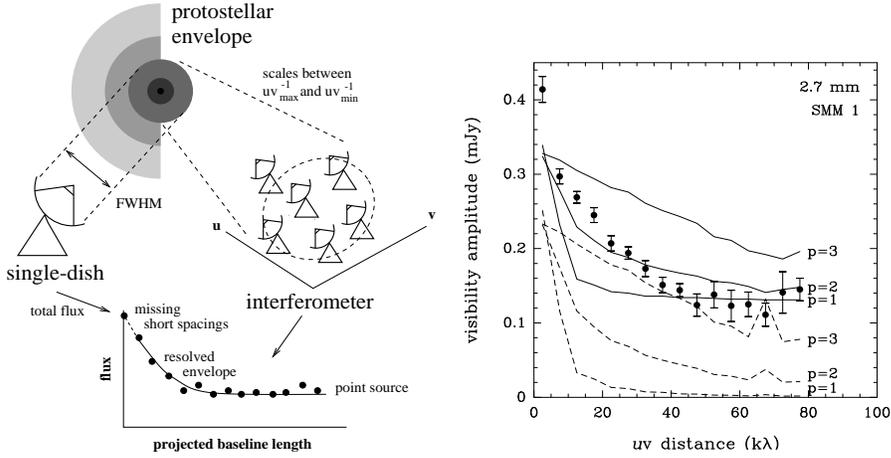,width=15cm,angle=270}
\end{center}
\caption{Left: Schematic illustration of single-dish versus
interferometer observations. Right:
Comparison of model visibility amplitudes at 2.7 mm to
observations of SMM~1. The observed vector-averaged amplitudes are
indicated by the filled symbols and their 1$\sigma$ error bars. The
dashed lines are models without a central point source, and density
power-law slopes of $-1.0$ ({\it lower curve in each panel\/}), $-2.0$
({\it middle curve\/}), and $-3.0$ ({\it upper curve\/}). Solid lines
are models with a point source flux of 0.13 Jy at 2.7 mm and a
spectral slope of 2.0.}
\label{fig-13}
\end{figure}

\section{Examples}

\subsection{An intermediate-mass class~0 YSO in Serpens}

Recent work by Hogerheijde et al.\ (1999) of the class~0 YSO Serpens
SMM~1 (FIRS~1) illustrates the use of molecular-line and continuum
observations with single-dish and aperture-synthesis instruments to
constrain the physical and chemical conditions in the envelope of a
low- to intermediate-mass protostar.  The Serpens molecular cloud
($d\approx 400$ pc) is in the process of forming a loosely bound
cluster of more than 50 stars (Eiroa \& Casali 1992). SMM~1 is one of
the submillimeter-continuum class~0 YSOs identified by Casali et al.\
(1993) with $L_{\rm bol}=77$ $L_\odot$ and $M_\star= 0.7$--$3.9$
$M_\odot$.  Table~\ref{tab-comp} gives an overview of the different
physical components traced by the various observations.

The continuum observations are useful to constrain both the mass and
the density structure of the envelope. The total mass is derived from
the single dish data, whereas the density variation of the envelope is
best obtained from the spatially resolved emission in the
interferometer. The method is illustrated on the left-hand side of
Figure~\ref{fig-13}. The dependence of the flux on projected
baseline length for SMM~1 (right hand side of Figure~\ref{fig-13})
indicates a power-law density distribution; different prescriptions
like, e.g., a Gaussian distribution, are excluded by the data. A
power-law index of $-2.0\pm 0.5$ gives a good fit to the observations,
which agrees well with models for protostellar collapse (e.g., Shu
1977). The fluxes on long baselines suggest the presence of extra,
unresolved emission, probably associated with the inner few hundred AU
of the envelope where the temperature is likely to exceed the adopted
$T_{\rm dust} \propto r^{-0.4}$ distribution. The absolute temperature
scale is $\sim$27 K at a characteristic radius of 1000 AU, constrained
by the spectral energy distribution over millimeter to far-infrared
wavelengths. The total envelope mass is 8.7 $M_\odot$, with an
uncertainty of a factor of 2--3 due to the adopted dust emissivity
(Ossenkopf \& Henning 1994). The fact that the inferred envelope mass
is larger than the stellar mass confirms SMM~1's nature as a class~0
YSO.

\begin{figure}[tb]
\vspace{-1.0cm}
\begin{center}
\leavevmode
\psfig{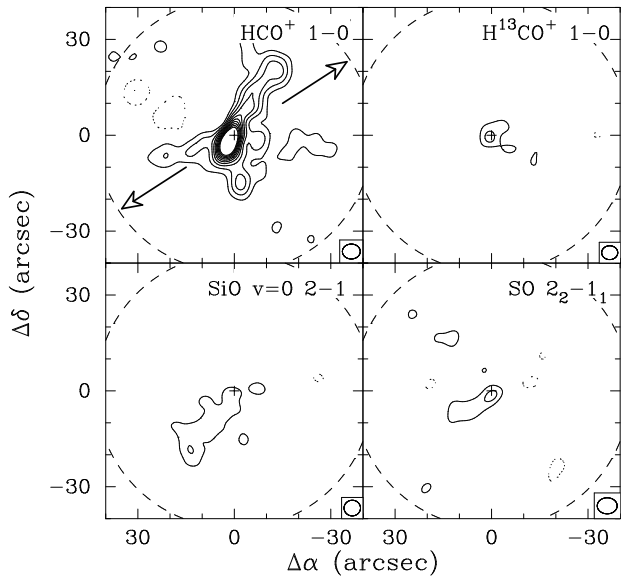}
\end{center}
\vspace{-0.5cm}
\caption{Cleaned, naturally weighted images of molecular line emission
observed with OVRO toward SMM~1. Contours are drawn at 3$\sigma$
intervals, of 1 K~km~s$^{-1}$ for HCO$^+$ and SO, 8 K~km~s$^{-1}$ for
H$^{13}$CO$^+$, and 4 K~km~s$^{-1}$ for SiO.The synthesized beam size
is indicated in each panel. The dashed circle shows the primary beam
size. The arrows in the HCO$^+$ panel indicate the position angle of
the radio jet from Rodr\'{\i}guez et~al.\ (1989) (from: Hogerheijde
et al.\ 1999).}
\label{fig-14}
\end{figure}

Using this description for the physical structure of the envelope, the
molecular abundances can be determined using models of the line
emission. Hogerheijde et al.\ (1999) adopt a Monte-Carlo technique to
solve the radiative transfer and the molecular excitation throughout
the envelope. In addition to the physical conditions, the molecular
line calculations require knowledge of the velocity structure.
Systematic velocity fields like infall and outflow can be included,
but in the simplest analysis a turbulent line width of $\sim$1.4
km~s$^{-1}$ independent of radius is adopted, based on
optically thin C$^{17}$O 3--2 and H$^{13}$CO$^+$
3--2 lines. The neglect of systematic velocity fields influences the
results for optically thick lines like $^{12}$CO and HCO$^+$. Note
that the lines themselves can in principle also constrain the density
structure (e.g., Hogerheijde et al. 1997), but are less accurate for
low- to intermediate-mass YSOs because of lack of spatial resolution
in the single-dish data and of signal-to-noise in the interferometer
observations.

The envelope mass can also be constrained from the optically thin CO
emission. If the `standard' dark cloud CO/H$_2$ abundance of $10^{-4}$
is adopted, the model overestimates the emission in the C$^{18}$O 2--1
and C$^{17}$O 3--2 lines by a factor of 2.  Apart from adjusting the
CO abundance, a gas kinetic temperature of 0.6 times the dust
temperature reconciles this discrepancy; $T_{\rm kin} < T_{\rm dust}$
is expected because of imperfect thermal coupling between the dust and
the gas, and line cooling of the gas. Alternatively, CO can freeze out
onto dust grains when the temperature drops below its sublimation
temperature of $\sim$20 K (see Table~\ref{tab-ice}), lowering the
emission. A CO depletion of a factor of 6 in the cold outer layers of
the envelope provides a good fit to the C$^{17}$O 3--2 and C$^{18}$O
2--1 lines. These lines are particularly good tracers of depletion
below 20 K, since the 2--1 lines, tracing $T_{\rm kin}\sim 16$ K gas,
are stronger affected than the 3--2 lines, tracing $T_{\rm kin}\sim 30$
K gas. The inferred depletion of a factor of $\sim$6 for SMM~1 is
smaller than found toward other class~0 envelopes (e.g., 10--20 for
NGC~1333 IRAS~4, Blake et al.\ 1995), suggesting that the evolutionary
phase characterized by large depletion factors may be short-lived, or
that local environment (e.g., heating of the envelope from the
outside) influences the depletion.

Most of the observed submillimeter transitions of other molecules
trace material in excess of $\sim 30$ K, and the derived abundances do
not depend critically on the exact value of $T_{\rm kin}$ or possible
depletion below 20 K. Table~\ref{tab-abun} lists the inferred
abundances for the parameters of the envelope derived from the dust
emission. Compared to other well-studied class~0 YSOs such as IRAS
16293$-$2422 (van Dishoeck et al.\ 1995), the abundances of HCO$^+$,
HCN, and H$_2$CO are very similar, while C$_3$H$_2$, CN, and HC$_3$N
are enhanced by an order of magnitude in SMM~1. The values for SiO and
SO are lower by a factor of 10, most likely because the $14''$--$19''$
single-dish beam of the SMM~1 observations does not include the tip of
the outflow where these species peak (see Figure~\ref{fig-14}).

Many but not all lines are well fit by this model.  High-excitation
lines like $^{13}$CO 6--5 and H$_2$CO $3_{22}$--$2_{21}$ indicate the
presence of an additional amount of warm, $\sim 100$ K, material. The
estimated column density of this gas amounts to less than 1\% of the
envelope mass, but dominates in emission of these highly excited
lines. It is likely that this material corresponds to the inner few
hundred AU of the envelope, where the temperature may exceed the
adopted distribution.  Some of the emission in the other lines may
originate in this component as well. Table~\ref{tab-abun} includes the
abundances derived under the assumption that all emission originates
in this warm gas. These values are larger by typically an order of
magnitude, illustrating that care needs to be taken to separate these
two contributions.

\begin{table}
\begin{center}
\caption{Derived molecular abundances in the envelope of Serpens SMM~1$^a$}
\label{tab-abun}
\begin{tabular}{lrrr}
\hline
Species & Envelope & Warm gas$^b$ & IRAS 16293$-$2442$^c$ \\
\hline
$^{12}$CO & $\equiv 1(-4)$ & $\equiv 1(-4)$ & $\ldots$ \\
HCO$^+$ & $1(-9)$ & $2(-8)$ & $2(-9)$ \\
HCN & $2(-9)$ & $5(-8)$ & $2(-9)$ \\
H$_2$CO & $8(-10)$ & $9(-9)$ & $7(-10)$ \\
C$_3$H$_2$ & $2(-10)$ & $3(-9)$ & $4(-11)$ \\
CN & $5(-9)$ & $3(-8)$ & $1(-10)$ \\
HC$_3$N & $2(-10)$ & $9(-10)$ & $3(-11)$ \\
SiO & $1(-11)$& $1(-10)$ & $1(-10)$ \\
SO & $2(-10)$ & $2(-9)$ & $4(-9)$ \\
\hline
\end{tabular}
\end{center}

$^a$ From Hogerheijde et al.\ (1999). \\

\vspace{-0.3cm}

$^b$ Abundances derived under the assumption that {\it all\/} emission 
originates in warm 100 K gas. \\

\vspace{-0.3cm}

$^c$ From van Dishoeck et al.\ (1995). \\

\end{table}

Figure~\ref{fig-14} shows the OVRO images of the low-excitation
lines of various molecules.  The emission in the line wings of
optically thick lines of CO, HCO$^+$ and HCN is clearly associated
with the bipolar outflow, as is that of SiO 2--1. HCO$^+$ outlines the
outflow cavity, possibly tracing the slow entrainment of material into
the flow. SiO coincides with the axis of the outflow, likely revealing
destruction of dust particles by the direct impact of the outflow on
ambient material. To derive abundances from these interferometer
observations, the modeling has to include spatial filtering. For
example, only 0.25 $M_\odot$ is traced by the optically thin C$^{18}$O
1--0 interferometer observations, but if the resolving-out of extended
emission is taken into account, this value is consistent with the
total envelope mass of 8.7 $M_\odot$. Firmer conclusions about
abundance variations on small scales requires more realistic, $>$1D
models of the physical structure, including a correct treatment of the
velocity field.


\begin{figure}[tb]
\vspace{-0.5cm}
\begin{center}
\leavevmode
\psfig{figure=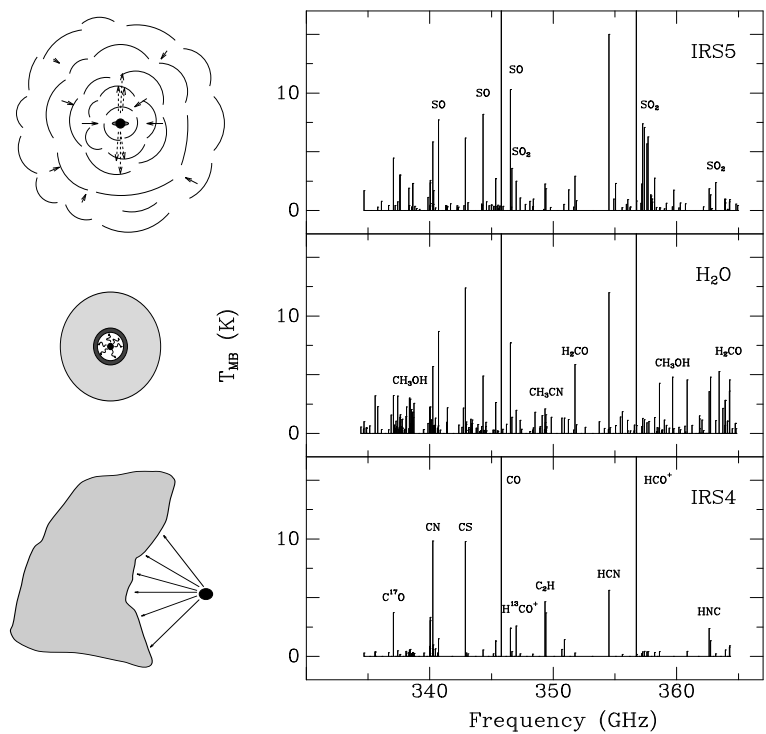,height=13.0cm,angle=-90}
\end{center}
\vspace{-1.5cm}
\caption{Summary of the JCMT 335--365 GHz line survey of three massive
star--forming regions in the W~3 molecular cloud. The spectra were
constructed from the observed line intensities obtained in
double--side-band mode.  Strong lines in common in the three spectra
are labelled in the W~3 IRS4 spectrum only.  Large physical and
chemical differences are found between the three regions, which are
attributed to different evolutionary stages  (Based on Helmich \& van
Dishoeck 1997).  }
\label{fig-15}
\end{figure}

\subsection{The W~3 massive star-forming region}

The W~3 massive star-forming region at $\sim 2.3$ kpc provides an
excellent opportunity to study the chemistry of massive YSOs
($L\approx 10^5 L_{\odot}$) at different evolutionary stages
originating from the same parent cloud.  Helmich et al.\ (1994) and
Helmich \& van Dishoeck (1997) performed an unbiased single-dish
submillimeter spectral survey of three YSOs at $15''$ (0.15 pc)
resolution: IRS5, IRS4 and W~3(H$_2$O). Although such observations
lack the spatial resolution of the data on nearby low-mass objects
discussed above, they provide a useful global picture of the
chemistry.  In addition, the envelope mass associated with these
objects is significantly larger than for low-mass YSOs, so that even
minor species can be detected.

Figure~\ref{fig-15} summarizes the spectra toward the three objects in
the 345 GHz atmospheric window. Clear physical and chemical
differences are found between the three sources, in spite of their
similar luminosities.  The beam-averaged densities are at least $10^6$
cm$^{-3}$ and the temperatures range from $\sim 55$~K for IRS4 to at
least 220~K for W~3(H$_2$O).  Toward W~3(IRS5), silicon- and
sulfur-bearing molecules such as SiO and SO$_2$ are prominent. This
source has a powerful, massive outflow and is presumably the youngest
of the three sources, showing also strong ice absorption features in
its infrared spectra. Toward W~3(H$_2$O), discovered by Turner \&
Welch (1984), organic molecules like CH$_3$OH, CH$_3$OCH$_3$, and
CH$_3$OCHO are at least one order of magnitude more abundant,
indicating that it is well into the hot core phase. Finally, only
simple molecules and radicals are found toward IRS4.  One possible
interpretation is that this object is more evolved and has already
broken free from the parent cloud, setting up an H II region and PDR
on the back side of the cloud.

More detailed physical models, which constrained by multi-line
observations of high-dipole molecules like CS, H$_2$CO, HCN and their
isotopes,   are needed to infer
reliable abundances. 
Recent analyses of the massive YSO GL~2591 suggest a
power-law density distribution with an index of $-1.0$ to $-1.5$,
somewhat shallower than found for low-mass YSOs such as SMM~1 (Carr et
al.\ 1995, van der Tak et al.\ 1999).

\section{Chemistry in circumstellar disks}

Most of this chapter has focused on the chemistry in the envelopes
of deeply-embedded YSOs. Young
stars with ages $1-5$ Myr have dispersed their envelopes
but are still surrounded by circumstellar disks with masses of
$\sim 10^{-3}-10^{-1}$ M$_{\odot}$ and sizes of $\sim$100~AU, comparable to
those inferred for our primitive solar nebula (see Beckwith \& Sargent 
1996 for a
review). Imaging of the gas in disks around T~Tauri and Herbig Ae stars
has so far been limited to CO and its isotopes, for reasons of
sensitivity. Observations of other molecules are only
just beginning. Single-dish surveys of various molecules have been
carried out for a few objects, in particular DM~Tau and GG~Tau
(Dutrey et al.\ 1997), TY~Hya (Kastner et al.\ 1997) and LkCa~15, MWC~480
and HD~163296 (Thi et al., in preparation). Millimeter
interferometers are now also capable of imaging the more abundant
molecules in disks, showing interesting morphological
differences (Qi et al., in preparation).

The detection of species like HCO$^+$, CN and HNC indicates that both
ion-molecule, photon-dominated and gas-grain chemistry play a role. The amount
of depletion is high because of the high densities (typically $10^6-10^9$
cm$^{-3}$ in the outer disk) and appears to be species-dependent, with the
most volatile molecules remaining in the gas furthest from the central star.
Chemical models which couple the gas-dust chemistry with the dynamical
evolution as the material is transported inwards have been developed recently
by Aikawa et al.\ (1997, 1998) and Willacy et al.\ (1998). At small disk
radii, irradiation from the central star and stellar X-rays may also affect
the chemistry. These studies are important first steps to determine the
connection between interstellar and solar nebula processes in the formation of
icy planetesimals such as comets and Kuiper-Belt objects.

\section{Concluding remarks}

Significant progress has been made in the last decade on models and
observations of the chemistry in star-forming regions. New
(sub-)millimeter data at higher angular resolution and frequencies
allow the dense and warm components of the YSO envelopes to be probed
directly. Ground- and spaced-based infrared observations provide the
first complete inventory of solid-state species along the line of
sight and put constraints on gas/solid ratios of abundant species. The
large variations in observed abundances illustrate that the chemistry
clearly responds to the enormous changes in temperature and density
during star-formation. Molecules freeze out onto grains in the cold
pre-stellar cores and outer envelopes surrounding YSOs, where
grain-surface chemistry produces new species. They are returned to the
gas by heating due to radiation from the young stars and by shocks due
to outflows impacting on the envelope. A rich chemistry can ensue in
the hot gas, leading to abundant complex organic molecules and driving
much of the available oxygen into water. Thus, these molecules provide
important chemical and temporal diagnostics of the YSO environment.

The derivation of abundances from spectral line data is far from
simple, however, and invariably includes assumptions about the
underlying source structure.  This physical structure can be
constrained by analyzing spatially resolved continuum data, as
illustrated for the case of Serpens SMM~1, and by using line ratios of
appropriate molecules to infer temperatures and densities.  Different
approaches, ranging from the ``homogeneous'' to the ``detailed''
method, offer optimal strategies for different cases.  Interferometric
data require special care, because the instrument is a spatial filter
which acts differently on lines tracing different physical conditions.

The development of comprehensive gas--grain models has made it
possible to put the observational results into a coherent framework of
the evolution of gas and solid species throughout the formation
of protostars to their incorporation into circumstellar disks and
eventually comets and planetesimals.  Such models are rapidly progressing
beyond the (pseudo) time-dependent method, and attempts are being made
to couple the chemistry with (multi-)dimensional hydrodynamics.
Accurate information on the basic molecular processes
entering such models remains a prerequisite.

Future instrumentation, such as the next generation
submillimeter arrays and air- or space-borne infrared telescopes
such as SOFIA, SIRTF, FIRST and NGST, will allow these studies to be
extended to much higher sensitivity and/or 
smaller scales of only a few AU. Together they should
provide a much clearer view on the connection between matter found in
interstellar clouds and in new planetary systems, and on the origin of
organic matter found in primitive objects in our own solar system.

\acknowledgements 
The authors are grateful to many colleagues for discussions and
preprints of their work.  They are indebted to the Leids
Kerkhoven-Bosscha Fonds and the organizers for financial
support.  Research on Astrochemistry in Leiden is supported by the
Netherlands Organization for Scientific Research (NWO).
MRH is supported by the Miller Institute for Basic Research in Science.


\begin{thebibliography}{}


\bibitem[]{} Aikawa, Y., Umebayashi, T., Nakano, T., Miyama, S.M. (1997),
{\it ApJ\/}, {\bf 486}, p.~L51

\bibitem[]{} Aikawa, Y., Umebayashi, T., Nakano, T., Miyama, S.M. (1998),
{\it Faraday Disc.}, {\bf 109}, p.~281

\bibitem[]{} Bachiller, R. (1997), in {\it Molecules in Astrophysics:
Probes and Processes\/}, IAU Symposium 178, ed.\ E.F. van Dishoeck.
Kluwer Academic Publishers, Dordrecht, p.~103

\bibitem[]{} Bachiller, R. and Per\'ez Guti\'errez, M. (1997), {\it
ApJ\/}, {\bf 487}, p.~L93

\bibitem[]{} Basu, S. and Mouschovias, T.C. (1994), {\it ApJ\/}, {\bf
432}, p.~720

\bibitem[]{} Beckwith, S.V.W. and Sargent, A.I. (1996), {\it Nature},
{\bf 383}, p.~139

\bibitem[]{} Benson, P.J., Caselli, P. and Myers, P.C. (1998), {\it
ApJ\/}, {\bf 506}, p.~743

\bibitem[]{} Bergin, E.A. and Langer, W.D. (1997), {\it ApJ\/}, {\bf
486}, p.~316

\bibitem[]{} Bergin, E.A., Goldsmith, P.F., Snell, R.L. and Langer,
W.D. (1997), {\it ApJ\/}, {\bf 482}, p.~285

\bibitem[]{} Bergin, E.A., Plume, R., Williams, J.P. and Myers, P.C.
(1999), {\it ApJ\/}, in press

\bibitem[]{} Bernes, C. (1979), {\it A\&A\/}, {\bf 73}, p.~67

\bibitem[]{} Bernstein, M.P., Sandford, S.A., Allamandola, L.J.,
Chang, S. and Scharberg, M.A. (1995), {\it ApJ\/}, {\bf 454}, p.~327

\bibitem[]{} Blake, D., Allamandola, L., Sandford, S., 
Hudgins, D., Freund, F. (1991), {\it Science}, {\bf 254}, pp. 548

\bibitem[]{} Blake, G.A. (1997), in {\it Molecules in Astrophysics:
Probes and Processes}, IAU Symposium 178, ed.\ E.F. van Dishoeck,
Kluwer Academic Publishers, Dordrecht, p.~31

\bibitem[]{} Blake, G.A., Sandell, G., van Dishoeck, E.F., Groesbeck,
T.D., Mundy, L.G. and Aspin, C. (1995), {\it ApJ\/}, {\bf 441},
p.~689

\bibitem[]{} Blake, G.A., Sutton, E.C., Masson, C.R. and Phillips,
T.G. (1987), {\it ApJ\/}, {\bf 315}, p.~621

\bibitem[]{} Boogert, A.C.A., Helmich, F.P., van Dishoeck, E.F.,
Schutte, W.A., Tielens, A.G.G.M. and Whittet, D.C.B. (1998), {\it
A\&A\/}, {\bf 336}, p.~352

\bibitem[]{} Boogert, A.C.A., Ehrenfreund, P., Gerakines, P.,
et al. (1999), {\it A\&A}, in press

\bibitem[]{} Campbell, M.F., Butner, H.M., Harvey, P.M., Evans, N.J.,
Campbell, M.B. and Sabbey C.N. (1995), {\it ApJ\/}, {\bf 454}, p.~831

\bibitem[]{} Carr, J.S., Evans, N.J., Lacy, J.L. and Zhou, S. (1995),
{\it ApJ\/}, {\bf 450}, p.~667

\bibitem[]{} Casali, M.M., Eiroa, C. and Duncan, W.D. (1993), {\it
A\&A\/}, {\bf 275}, p.~195

\bibitem[]{}Caselli, P., Hasegawa, T.I. and Herbst, E. (1993), {\it
ApJ\/}, {\bf 408}, p.~548

\bibitem[]{}Caselli, P., Hasegawa T.I. and Herbst, E. (1998), {\it
ApJ\/}, {\bf 495}, p.~309

\bibitem[]{}Ceccarelli, C., Hollenbach, D.J. and Tielens, A.G.G.M.
(1996), {\it ApJ\/}, {\bf 471}, p.~400

\bibitem[]{} Ceccarelli, C., Castets, A., Loinard, L., Caux, E. and
Tielens, A.G.G.M. (1998), {\it A\&A\/}, {\bf 338}, p.~L43

\bibitem[]{} Cernicharo, J., Gonz\'alez-Alfonso, E., Alcolea, J.,
Bachiller, R. and John, D. (1994), {\it ApJ\/}, {\bf 432}, p.~L59

\bibitem[]{} Charnley, S.B. (1997), {\it ApJ\/}, {\bf 481}, p.~396

\bibitem[]{} Charnley, S.B, Kress, M.E., Tielens, A.G.G.M. and Millar,
T.J. (1995), {\it ApJ\/}, {\bf 448}, p.~232

\bibitem[]{} Charnley, S.B., Tielens, A.G.G.M. and Millar,
T.J. (1992), {\it ApJ\/}, {\bf 399}, p.~L71

\bibitem[]{} Chiar, J.E., Adamson, A.J., Kerr, T.H. and Whittet,
D.C.B. (1995), {\it ApJ\/}, {\bf 455}, p.~234

\bibitem[]{} Chiar, J.E., Gerakines, P.A., Whittet, D.C.B., Pendleton,
Y.J., Tielens, A.G.G.M., Adamson, A.J. and Boogert, A.C.A. (1998),
{\it ApJ\/}, {\bf 498}, p.~716

\bibitem[]{} Dalgarno, A. (1987), in {\it Physical Processes in
Interstellar Clouds}, eds. G. Morfill and M.S. Scholer, D. Reidel,
Dordrecht, p.~219

\bibitem[]{}Dalgarno, A. (1994), {\it Adv. At. Mol. \& Opt. Phys.\/},
{\bf 32}, p.~57

\bibitem[]{}Dartois, E., d'Hendecourt, L., Boulanger, F., Jourdain de Muizon,
M., Breitfellner, M., Puget, J.-L., and Habing, H.J. (1998),
{\it A\&A\/}, {\bf 331}, p.~651

\bibitem[]{} de Boisanger, C., Helmich, F.P. and van Dishoeck,
E.F. (1996), {\it A\&A\/}, {\bf 310}, p.~315

\bibitem[]{}d'Hendecourt, L., Jourdain de Muizon, M., Dartois, E.,
Breitfellner, M., Ehrenfreund, P., Benit, J., Boulanger, F., Puget,
J.L. and Habing H.J. (1996), {\it A\&A\/}, {\bf 315}, p.~L365

\bibitem[]{} Doty, S.D. and Neufeld, D.A. (1997), {\it ApJ\/}, {\bf
489}, p.~122

\bibitem[]{} Draine, B.T. (1978), {\it ApJS\/}, {\bf 36}, p.~595

\bibitem[]{} Draine, B.T. and McKee, C.F. (1993), {\it ARA\&A\/}, {\bf
31}, p.~373

\bibitem[]{} Dutrey, A., Guilloteau, S. and Gu\'elin, M. (1997),
{\it A\&A\/}, {\bf 317}, p.~L55

\bibitem[]{} Eiroa, C. and Casali, M.M. (1992), {\it A\&A\/}, {\bf
262}, p.~468

\bibitem[]{} Ehrenfreund, P., d'Hendecourt, L., Dartois, E., Jourdain
de Muizon, M., Breitfellner, M., Puget, J.L. and Habing, H.J. (1997),
{\it Icarus\/}, {\bf 130}, p.~1

\bibitem[]{} Ehrenfreund, P., Dartois, E., Demyk, K. and d'Hendecourt,
L. (1998), {\it A\&A\/}, {\bf 339}, p.~L17

\bibitem[]{} Evans, N.J. (1980), in {\it Interstellar Molecules},
Proc.\ IAU 87, ed.\ B.H.\ Andrew, D. Reidel, Dordrecht, p.~1

\bibitem[]{} Evans, N.J., Lacy, J.H. and Carr, J.S. (1991), {\it
ApJ\/}, {\bf 383}, p.~674


\bibitem[]{} Flower, D. (1990). Molecular Collisions in the Interstellar
Medium, Cambridge University Press

\bibitem[]{} Frerking, M.A., Langer, W.D. and Wilson, R.W. (1982),
{\it ApJ\/}, {\bf 262}, p.~590

\bibitem[]{}Geballe, T.R. and Oka, T. (1996), {\it Nature\/}, {\bf
384}, p.~334

\bibitem[]{}Gensheimer, P.D., Mauersberger, R. and Wilson,
T.L. (1996), {\it A\&A\/}, {\bf 314}, p.~281

\bibitem[]{}Genzel, R. (1992), in {\it The Galactic Interstellar
Medium}, eds. D. Pfenniger and P. Bartholdi, Saas Fee Advanced Course
21, Springer, Berlin, p.~275

\bibitem[]{}Gerakines, P.A., Schutte, W.A. and Ehrenfreund, P. (1996),
{\it A\&A\/}, {\bf 312}, p.~289

\bibitem[]{}Gerakines, P.A., Whittet, D.C.B., Ehrenfreund, P.\
 et al. (1999), {\it ApJ}, in press

\bibitem[]{}Gibb, E., Whittet, D.C.B., et al.\ (1999), in preparation

\bibitem[]{}Gredel, R., van Dishoeck, E.F. and Black, J.H. (1993),
{\it A\&A\/}, {\bf 269}, p.~477

\bibitem[]{}Gredel, R., Lepp, S., Dalgarno, A. and Herbst, E. (1989),
{\it ApJ\/}, {\bf 347}, p.~289

\bibitem[]{} Green, S. (1975), in Atomic and molecular physics and the
interstellar matter, ed. R. Balian et al., Elsevier North Holland, Amsterdam,
p. 83



\bibitem[]{} Habing, H.J. (1988), in {\it Millimetre and Submillimetre
Astronomy}, ed.\ R.D.\ Wolstencroft
and W.B. Burton , Kluwer Academic Publishers, Dordrecht, p.~207


\bibitem[]{} Hartquist T.W., Caselli, P., Rawlings, J.M.C.,
Ruffle, D.P., Williams, D.A. (1998), in {\it The Molecular
Astrophysics of Stars and Galaxies}, ed.\ T.W.\ Hartquist and D.A.\
Williams, Oxford, Oxford University, p.~101

\bibitem[]{}Hasegawa, T.I. and Herbst, E. (1993), {\it MNRAS\/}, {\bf
261}, p.~83

\bibitem[]{}Hatchell, J., Thompson, M.A., Millar, T.J. and Macdonald,
G.H. (1998), {\it A\&AS\/}, {\bf 133}, 29

\bibitem[]{}Helmich, F.P. and van Dishoeck, E.F. (1997), {\it
A\&AS\/}, {\bf 124}, p.~205

\bibitem[]{}Helmich, F.P., Jansen, D.J., de Graauw, Th., Groesbeck,
T.D. and van Dishoeck, E.F. (1994), {\it A\&A\/}, {\bf 283}, p.~626

\bibitem[]{}Herbst, E. (1993), in {\it Dust and Chemistry in
Astronomy}, eds. T.J. Millar and D.A. Williams, IOP, Bristol, p.~183

\bibitem[]{}Herbst, E. (1995), {\it Ann. Rev. Phys. Chem.\/}, {\bf
46}, p.~27

\bibitem[]{}Herbst, E. and Leung, C.M. (1986), ApJ, {\bf 310}, 378.


\bibitem[]{}Herbst, E. and Klemperer, W. (1973), {\it ApJ}, {\bf 185},
p.~505

\bibitem[]{} Hogerheijde, M.R., Jansen, D.J. and van Dishoeck,
E.F. (1995), {\it A\&A\/}, {\bf 294}, p.~792

\bibitem[]{} Hogerheijde, M.R., van Dishoeck, E.F., Blake, G.A. and
van Langevelde, H.J. (1997), {\it ApJ\/}, {\bf 489}, p.~293

\bibitem[]{}Hogerheijde, M.R., van Dishoeck, E.F., Blake, G.A. and van
Langevelde, H.J. (1998), {\it ApJ\/}, {\bf 502}, p.~315

\bibitem[]{}Hogerheijde, M.R., van Dishoeck, E.F., Salverda, J.M. and
Blake, G.A. (1999), {\it ApJ\/}, in press

\bibitem[]{}Hollenbach, D.J.\ (1997), in {\it Herbig-Haro Objects and
the Birth of Low-mass Stars, IAU Symposium 182}, ed.\ B.\ Reipurth \&
C.\ Bertout, Kluwer Academic Publishers, Dordrecht, p.~181

\bibitem[]{}Hollenbach, D.J. and Salpeter, E.E. (1971), {\it ApJ\/},
{\bf 163}, p.~155

\bibitem[]{}Hollenbach, D.J. and Tielens, A.G.G.M.\ (1997), {\it
ARA\&A\/}, {\bf 35}, p.179

\bibitem[]{} Hollenbach, D.J. and Tielens, A.G.G.M. (1999), {\it
Rev.\ Mod.\ Phys.}, in press

\bibitem[]{}Irvine, W.M.\ (1998), in {\it Origins of Life and
Evolution of the Biosphere}, {\bf 28}, p.~365

\bibitem[]{} Irvine W.M., Schloerb, F.P., Crovisier, J., Fegley, B., 
and Mumma, M.J.\ (1999), {\it Protostars \& Planets IV},
eds.\ V.G.\ Mannings and S.\ Russell, University of Arizona, in press

\bibitem[]{} Kaiser, R.I. and Roessler, K. (1997), {\it ApJ\/}, {\bf
475}, p.~487

\bibitem[]{} Kaiser, R.I., Stranges, D., Lee, Y.T. and Suits,
A.G. (1997), {\it ApJ\/}, {\bf 477}, p.~982

\bibitem[]{} Kastner, J.H., Zuckerman, B., Weintraub, D.A. and Forveille,
T. (1997), {\it Science}, {\bf 277}, p.~67

\bibitem[]{}Kaufman, M.J., Hollenbach, D.J. and Tielens, A.G.G.M.\
(1998), {\it ApJ\/}, {\bf 497}, p.~276

\bibitem[]{}Kramer, C., Alves, J., Lada, C.J., Lada, E.A., Sievers, A.,
Ungerechts, H. and Walmsley, C.M. (1999), {\it A\&A\/}, {\bf 329}, p.~L33

\bibitem[]{}Kuan, Y.J. and Snyder, L.E.\ (1996), {\it ApJ\/}, {\bf
470}, p.~981

\bibitem[]{}Kuiper, T.B.H., Langer, W.D. and Velusamy, T.\ (1996),
{\it ApJ\/}, {\bf 468}, p.~761

\bibitem[]{}Lacy, J.H., Carr, J.S., Evans, N.J., Baas, F., Achtermann,
J.M. and Arens, J.F. (1991), {\it ApJ\/}, {\bf 376}, p.~556

\bibitem[]{}Lacy, J.H., Evans, N.J., Achtermann, J.M., Bruce, D.E.,
Arens, J.F. and Carr, J.S. (1989), {\it ApJ\/}, {\bf 342}, p.~L43

\bibitem[]{}Lacy, J.H., Knacke, R., Geballe, T.R. and Tokunaga,
A.T. (1994), {\it ApJ\/}, {\bf 428}, p.~L69

\bibitem[]{}Lacy, J.H., Faraji, H., Sandford, S.A. and Allamandola,
L.J.\ (1998), {\it ApJ\/}, {\bf 501}, p.~L105

\bibitem[]{}Lada, C.J., Lada, E.A., Clemens, D. and Bally, J. (1994).
{\it ApJ\/}, {\bf 429}, p.~694

\bibitem[]{}Lahuis, F. and van Dishoeck, E.F.\ (1999), {\it A\&A\/},
to be submitted

\bibitem[]{} Langer, W.D., Velusamy, T., Kuiper, T.B.H., Peng, R.,
McCarthy, M.C., Travers, M.J., Kovacs, A., Gottlieb, C.A. and
Thaddeus, P. (1997), {\it ApJ\/}, {\bf 480}, p.~L63

\bibitem[]{} Langer, W.D., van Dishoeck, E.F., Blake, G.A. et al. (1999), 
{\it Protostars \& Planets IV}, eds.\ V.G. Mannings and S. Russell,
University of Arizona, in press

\bibitem[]{} Larsson, M. (1997), {\it Ann. Rev. Phys. Chem.}, {\bf 48}, 151

\bibitem[]{} le Bourlot, J., Pineau des For\^ets, G., Roueff, E. and
Flower, D.R. (1995), {\it A\&A\/}, {\bf 302}, p.~870

\bibitem[]{}Lee, H.-H., Bettens, R.P.A. and Herbst, E. (1996a), {\it
A\&AS\/}, {\bf 119}, p.~111

\bibitem[]{}Lee, H.-H., Herbst, E., Pineau des For\^ets, G., Roueff,
E. and le Bourlot, J. (1996b), {\it A\&A\/}, {\bf 311}, p.~690

\bibitem[]{}Lee, H.-H., Roueff, E., Pineau des For\^ets, G., Shalabiea, O.,
Terzieva, R. and Herbst, E.\ (1998), {\it A\&A\/}, {\bf 334}, p.~1047

\bibitem[]{}Liseau, R., Ceccarelli, C., Larsson, B.\ et al.\ (1996),
{\it A\&A\/}, {\bf 315}, p.~L181

\bibitem[]{}Lucas, R. and Liszt, H. (1997), in {\it Molecules in
Astrophysics: Probes and Processes}, IAU Symposium 178, ed.\ E.F. van
Dishoeck, Kluwer Academic Publishers, Dordrecht, p.~421

\bibitem[]{}Macdonald, G.H., Gibb, A.G., Habing, R.J. and Millar,
T.J.\ (1996), {\it A\&AS\/}, {\bf 119}, p.~333

\bibitem[]{}Maloney, P.R., Hollenbach, D.J. and Tielens, A.G.G.M.\
(1996), {\it ApJ\/}, {\bf 466}, p.~561

\bibitem[]{} Marechal, P., Pagani, L., Langer, W.D. and Castets,
A. (1997), {\it A\&A\/}, {\bf 318}, p.~252

\bibitem[]{}McCall, B.J., Hinkle, K.H., Geballe, T.R. and Oka, T.\
(1998), {\it J.\ Chem\ Soc.\ Far.\ Disc.\/}, {\bf 109}, p.\ 267

\bibitem[]{}Melnick, G., Stauffer, J.R., Ashby, M.\ et al.\ (1998),
BAAS, 193, 72.01

\bibitem[]{}Meyer, D.M. (1997), in {\it Molecules in Astrophysics:
Probes and Processes}, IAU Symposium 178, ed.\ E.F. van Dishoeck,
Kluwer Academic Publishers, Dordrecht, p.~407

\bibitem[]{}Millar, T.J. (1990), in {\it Molecular Astrophysics ---A
volume honoring Alexander Dalgarno}, ed. T.W. Hartquist, Cambridge
University Press, p.~115

\bibitem[]{}Millar, T.J., Bennett, A. and Herbst, E. (1989), {\it
ApJ\/}, {\bf 340}, p.~906

\bibitem[]{}Millar, T.J., Farquhar, P.R.A. and Willacy, K. (1997a),
{\it A\&AS\/}, {\bf 121}, p.~139

\bibitem[]{}Millar, T.J., Macdonald, G.H. and Gibb, A.G.\ (1997b),
{\it A\&A\/}, {\bf 325}, p.~1163

\bibitem[]{}Millar, T.J. and Hatchell, J.\ (1998), {\it J.\ Chem\
Soc.\ Far.\ Disc.\/}, {\bf 109}, p.~15

\bibitem[]{} Mitchell, G.F., Maillard, J.-P., Allen, M., Beer, R. and
Belcourt, K. (1990), {\it ApJ\/}, {\bf 363}, p.~554

\bibitem[]{}Moore, M.H., Donn, B., Khanna, R. and A'Hearn, M.F.\
(1983), {\it Icarus\/}, {\bf 54}, p.~388

\bibitem[]{}Motte, F., Andr\'e, P. and Neri, R.\ (1998), {\it A\&A\/},
{\bf 336}, p.~150

\bibitem[]{}Mundy, L. and McMullin, J.P. (1997), in {\it Molecules in
Astrophysics: Probes and Processes}, IAU Symposium 178, ed.\ E.F. van
Dishoeck, Kluwer Academic Publishers, Dordrecht, p.~183

\bibitem[]{}Myers, P.C. and Benson, P.J. (1983), {\it ApJ\/}, {\bf
266}, p.~309

\bibitem[]{}Neufeld, D.A. and Dalgarno, A.\ (1989a), {\it ApJ\/}, {\bf
340}, p.~869

\bibitem[]{}Neufeld, D.A. and Dalgarno, A.\ (1989b), {\it ApJ\/}, {\bf
344}, p.~251

\bibitem[]{}Ohishi, M. (1997), in {\it Molecules in Astrophysics:
Probes and Processes}, IAU Symposium 178, ed.\ E.F. van Dishoeck,
Kluwer Academic Publishers, Dordrecht, p.~61

\bibitem[]{}Olano, C.A., Walmsley, C.M. and Wilson, T.L. (1988), {\it
A\&A\/}, {\bf 196}, p.~194

\bibitem[]{} Olofsson, G., Pagani, L., Tauber, J., et al. (1998), {\it
A\&A\/}, {\bf 339}, p.~81

\bibitem[]{} Ossenkopf, V. and Henning, T. (1994), {\it A\&A\/}, {\bf
291}, p.~943

\bibitem[]{} Osterbrock (1989), {\it Astrophysics of Gaseous Nebulae
and Active Galactic Nuclei\/}, University Science Books, Mill Valley

\bibitem[]{}Pineau des For\^ets, G. and Flower, D.R.\ (1997), in {\it
Molecules in Astrophysics: Probes and Processes}, IAU Symposium 178,
ed.\ E.F. van Dishoeck, Kluwer Academic Publishers, Dordrecht, p.~113

\bibitem[]{}Plume, R., Bergin, E.A., Williams, J.P. and Myers, P.C.\
(1998), {\it J.\ Chem\ Soc.\ Far.\ Disc.\/}, {\bf 109}, p.~47

\bibitem[]{}Prasad, S.S. and Huntress, W.T.\ (1980), {\it ApJS\/},
{\bf 43}, p.~1

\bibitem[]{}Pratap, P., Dickens, J.E., Snell, R.L.\ et al.\ (1997),
{\it ApJ\/}, {\bf 486}, p.~862

\bibitem[]{}Rawlings, J.M.C., Hartquist, T.W., Menten, K.M. and
Williams, D.A.  (1992), {\it MNRAS\/}, {\bf 255}, p.~471

\bibitem[]{}Roberge, W.G., Jones, D., Lepp, S. and Dalgarno,
A. (1991), {\it ApJS\/}, {\bf 77}, p.~287

\bibitem[]{} Rodr\'{\i}guez, L.F., Curiel, S., Moran, J., Mirabel,
I.F., Roth, M. and Garay, G. (1989), {\it ApJ\/}, {\bf 346}, p.~L85

\bibitem[]{} Sandford, S.A. and Allamandola, L.J. (1990), {\it
Icarus\/}, {\bf 87}, p.~188

\bibitem[]{} Sandford, S.A. and Allamandola, L.J. (1993), {\it ApJ\/},
{\bf 417}, p.~815

\bibitem[]{}Schilke, P., Walmsley, C.M., Pineau des For\^ets, G. and
Flower, D.R.\ (1997), {\it A\&A\/}, {\bf 321}, p.~293

\bibitem[]{}Schreyer, K., Helmich, F.P., van Dishoeck, E.F. and
Henning, T.\ (1997), {\it A\&A\/}, {\bf 326}, p.~347

\bibitem[]{} Schutte, W.A. (1996), in {\it The Cosmic Dust Connection},
ed.\ J.M. Greenberg, Kluwer Academic Publishers, Dordrecht, p.~1


\bibitem[]{} Schutte, W.A. (1999), in {\it Laboratory astrophysics
and space missions}, eds.\ P.\ Ehrenfreund et al., Kluwer Academic
Publishers, Dordrecht, p.\ 69

\bibitem[]{} Schutte, W.A., Tielens, A.G.G.M. and Allamandola,
L.J. (1993), {\it ApJ\/}, {\bf 415}, p.~397

\bibitem[]{}Shalabiea, O. and Greenberg, J.M. (1995), {\it A\&A\/},
{\bf 296}, p.~779

\bibitem[]{}Shalabiea, O.M., Caselli, P. and Herbst, E.\ (1998), {\it
ApJ\/}, {\bf 502}, p.~652

\bibitem[]{} Shu, F.H.\ (1977), {\it ApJ\/}, {\bf 214}, p.~488

\bibitem[]{} Shu, F.H., Adams, F.C., and Lizano, S. (1987), {\it
ARA\&A\/}, {\bf 25}, p.~23

\bibitem[]{}Smith, I.W.M. (1997), in {\it Molecules in Astrophysics:
Probes and Processes}, IAU Symposium 178, ed.\ E.F. van Dishoeck,
Kluwer Academic Publishers, Dordrecht, p.~253

\bibitem[]{} Sobolev, V.V. (1960), {\it Moving Envelopes of Stars\/},
Harvard University Press, Cambridge

\bibitem[]{}Sternberg, A., Yan, M. and Dalgarno, A.\ (1997), in {\it
Molecules in Astrophysics: Probes and Processes}, IAU Symposium 178,
ed.\ E.F. van Dishoeck, Kluwer Academic Publishers, Dordrecht, p.~141

\bibitem[]{} Sutton, E.C., Peng, R., Danchi, W.C., Jaminet, P.A.,
Sandell, G., and Russell, A.P.G. (1995), {\it ApJS\/}, {\bf 97},
p.~455

\bibitem[]{}Suzuki, H., Yamamoto, S., Ohishi, M. et al., (1992), {\it
ApJ\/}, {\bf 392}, p.~551

\bibitem[]{}Teixeira, T.C., Emerson, J.P. and Palumbo, M.E. (1998),
{\it A\&A\/}, {\bf 330}, p.~711

\bibitem[]{} Terzieva, R. and Herbst, E. (1998), {\it ApJ\/}, {\bf
501}, p.~207

\bibitem[]{}Tielens, A.G.G.M.\ (1983), {\it A\&A\/}, {\bf 119}, p.~177

\bibitem[]{}Tielens, A.G.G.M. and Allamandola, L.J. (1987), in {\it
Interstellar Processes}, eds. D. Hollenbach and H.A. Thronson,
D. Reidel, Dordrecht, p.~379

\bibitem[]{}Tielens, A.G.G.M. and Charnley, S.B.\ (1997), {\it Origin
of Life\/}, {\bf 27}, p.~23

\bibitem[]{}Tielens, A.G.G.M. and Hagen, W. (1982), {\it A\&A\/}, {\bf
114}, p.~245

\bibitem[]{}Tielens, A.G.G.M., Tokunaga, A.T., Geballe, T.R. and Baas,
F.\ (1991), {\it ApJ\/}, {\bf 381}, p.~181

\bibitem[]{}Tielens, A.G.G.M. and Whittet, D.C.B. (1997), in {\it
Molecules in Astrophysics: Probes and Processes}, IAU Symposium 178,
ed.\ E.F. van Dishoeck, Kluwer Academic Publishers, Dordrecht, p.~45

\bibitem[]{}Turner, B.E. (1996), {\it ApJ\/}, {\bf 468}, p.~694.

\bibitem[]{}Turner, B.E.\ (1998), {\it ApJ\/}, {\bf 495}, p.~804

\bibitem[]{}Turner, B.E., Lee, H.H. and Herbst, E.\ (1998), {\it
ApJS\/}, {\bf 115}, p.~91

\bibitem[]{}Turner, J.L. and Welch, W.J.\ (1984), {\it ApJ\/}, {\bf
287}, p.~L81

\bibitem[]{} Vandenbusssche, B., Ehrenfreund, P., Boogert, A.C.A.,
et al. (1999), {\it A\&A}, submitted

\bibitem[]{}van der Tak, F., van Dishoeck, E.F., Evans, N.J., Bakker, E. and
Blake, G.A.\ (1999), {\it ApJ\/}, submitted

\bibitem[]{}van Dishoeck, E.F. (1988), in {\it Millimetre and
Submillimetre Astronomy}, ed. R.D. Wolstencroft and W.B. Burton,
Kluwer Academic Publishers, Dordrecht, p.~117

\bibitem[]{}van Dishoeck, E.F.\ (1998a), in {\it The Molecular
Astrophysics of Stars and Galaxies}, ed.\ T.W.\ Hartquist and D.A.\
Williams, Oxford, Oxford University, p. 53

\bibitem[]{}van Dishoeck, E.F.\ (1998b), {\it Faraday
Disc.\/}, {\bf 109}, p.~31

\bibitem[]{} van Dishoeck, E.F. and Black, J.H. (1987), in {\it
Physical Processes in Interstellar Clouds}, eds. G. Morfill and
M.S. Scholer, D. Reidel, Dordrecht, p.~241

\bibitem[]{}van Dishoeck, E.F. and Black, J.H. (1988), {\it ApJ\/},
{\bf 334}, p.~771

\bibitem[]{}van Dishoeck, E.F. and Blake, G.A.\ (1998), {\it ARA\&A},
{\bf 36}, p.~317

\bibitem[]{}van Dishoeck, E.F., Blake, G.A., Draine, B.T., Lunine,
J.I. (1993), in {\it Protostars and Planets III}, eds. E.H. Levy and
J.I. Lunine, Univ. of Arizona, p. 163

\bibitem[]{} van Dishoeck, E.F., Blake, G.A., Jansen, D.J. and
Groesbeck, T.D. (1995), {\it ApJ\/}, {\bf 447}, p.~760

\bibitem[]{}van Dishoeck, E.F., Helmich, F.P., de Graauw, Th. et
al. (1996), {\it A\&A\/}, {\bf 315}, p.~L349

\bibitem[]{}van Dishoeck, E.F., Wright, C.M., Cernicharo, J., et
al. (1998), {\it ApJ\/}, {\bf 502}, p.~L173

\bibitem[]{} van Dishoeck, E.F., Helmich, F.P., Schutte, W.A., et al.\
1998, in {\it Star Formation with the Infrared Space Observatory},
eds.\ J. Yun and R.\ Liseau, ASP vol. 132, p. 54


\bibitem[]{}Vejby-Christensen, L., Andersen, L.H., Heber, O., Kella,
D., Pedersen, H.B., Schmidt, H. and Zajfman, D. (1997), {\it ApJ\/},
{\bf 483}, p.~531

\bibitem[]{} Walmsley, C.M. (1987), in {\it Physical Processes in
Interstellar Clouds\/}, eds. G. Morfill and M.S. Scholer, D. Reidel,
Dordrecht, p.~161

\bibitem[]{} Walmsley, C.M. (1991), in {\it Fragmentation of molecular
clouds and star formation}, IAU Symposium 147, eds.\ E. Falgarone
et al., Kluwer Academic Publishers, Dordrecht, p.~161


\bibitem[]{}Walmsley, C.M. and Schilke, P. (1993), in {\it Dust and
Chemistry in Astronomy}, eds.\ T.J. Millar and D.A. Williams, IOP
Publishing, Bristol, p.~37

\bibitem[]{}Ward-Thompson, D., Scott, P.F., Hills, R.E. and Andr\'e,
P. (1994), {\it MNRAS\/}, {\bf 268}, p.~276

\bibitem[]{}Whittet, D.C.B. (1993), in {\it Dust and Chemistry in
Astronomy}, eds.\ T.J. Millar and D.A. Williams, IOP Publishing,
Bristol, p.~9.

\bibitem[]{}Whittet, D.C.B., Gerakines, P.A., Tielens, A.G.G.M.\ et
al.\ (1998), {\it ApJ\/}, {\bf 498}, p.~L159

\bibitem[]{}Whittet, D.C.B., Schutte, W.A., Tielens, A.G.G.M. et
al. (1996), {\it A\&A\/}, {\bf 315}, p.~L357


\bibitem[]{}Willacy, K.R., Rawlings, J.M.C. and Williams, D.A.\
(1994), {\it MNRAS\/}, {\bf 269}, p.~921

\bibitem[]{}Willacy, K.R., Klahr, H.H., Millar, T.J. and Henning, Th. (1998),
{\it A\&A\/}, {\bf 338}, p.~995

\bibitem[]{}Williams, D.A. (1993), in {\it Dust and Chemistry in
Astronomy}, eds.\ T.J. Millar and D.A. Williams, IOP Publishing,
Bristol, p.~143

\bibitem[]{}Williams, J.P., Bergin, E.A., Caselli, P., Myers, P.C. and
Plume, R.\ (1998), {\it ApJ\/}, {\bf 503}, p.~689

\bibitem[]{}Wootten, A. (1987), in {\it Astrochemistry}, eds.\
M.S. Vardya and S.P. Tarafdar, Kluwer Academic Publishers, Dordrecht, p.~311



\end{thebibliography}
\end{document}